\documentclass[amsmath,amssymb,superscriptaddress,aip,jcp,reprint,widetext]{revtex4-2}

\usepackage[utf8]{inputenc}
\usepackage{bm}
\usepackage{graphicx}
\usepackage{xcolor}
\usepackage{notes2bib}
\usepackage{siunitx}
\usepackage{xcolor}
\usepackage{hyperref}
\DeclareSIUnit\Molar{\textsc{m}}

\newcommand{\av}[1]{\left< #1 \right>}
\newcommand{\kbt}{k_{B}T}
\newcommand{\del}{\mathbf{\partial}}
\newcommand{\Fex}{\mathbf{F}_\text{ex}}
\newcommand{\trap}{\text{trap}}

\newcommand{\Fi}{\mathbf{F}_i}
\newcommand{\lr}{\text{lr}}
\newcommand{\eq}{\text{eq}}
\newcommand{\ex}{\text{ex}}

\newcommand{\sh}{\text{sh}}
\newcommand{\MSD}{\delta x^2}
\newcommand{\recoil}{\delta x(t)}

\begin{document}

\title{How are mobility and friction related in viscoelastic fluids?}
\author{Juliana Caspers}
\email{j.caspers@theorie.physik.uni-goettingen.de}
\affiliation{Institute for Theoretical Physics, Georg-August Universit\"{a}t G\"{o}ttingen, 37073 G\"{o}ttingen, Germany}
\author{Nikolas Ditz}
\affiliation{Fachbereich Physik, Universit\"{a}t Konstanz, 78457 Konstanz, Germany}
\author{Karthika Krishna Kumar}
\affiliation{Fachbereich Physik, Universit\"{a}t Konstanz, 78457 Konstanz, Germany}
\author{F\'elix Ginot}
\affiliation{Fachbereich Physik, Universit\"{a}t Konstanz, 78457 Konstanz, Germany}
\author{Clemens Bechinger}
\affiliation{Fachbereich Physik, Universit\"{a}t Konstanz, 78457 Konstanz, Germany}
\author{Matthias Fuchs}
\affiliation{Fachbereich Physik, Universit\"{a}t Konstanz, 78457 Konstanz, Germany}
\author{Matthias Krüger}
\affiliation{Institute for Theoretical Physics, Georg-August Universit\"{a}t G\"{o}ttingen, 37073 G\"{o}ttingen, Germany}

\date{\today}

\begin{abstract}
The motion of a colloidal probe in a viscoelastic fluid is described by friction or mobility, depending on whether the probe is moving with a velocity or feeling a force. 
While the Einstein relation describes an inverse relationship valid for Newtonian solvents, both concepts are generalized to time-dependent memory kernels in viscoelastic fluids.
We theoretically and experimentally investigate their relation by considering two observables: the recoil after releasing a probe that was moved through the fluid and the equilibrium mean squared displacement (MSD).
Applying concepts of linear response theory, we generalize Einstein's relation and thereby relate recoil and MSD, which both provide access to the mobility kernel.
With increasing concentration, however, MSD and recoil show distinct behaviors, rooted in different behaviors of the two kernels.
Using two theoretical models, a linear two-bath particle model and hard spheres treated by mode-coupling theory, we find a Volterra relation between the two kernels, explaining differing timescales in friction and mobility kernels under variation of concentration.
\end{abstract}

\maketitle


\section{Introduction}

Observing the  Brownian motion of colloidal probe particles can be used to investigate complex fluids,  soft materials or biological tissues \cite{larson_structure_1999,squires2010fluid,puertas_microrheology_2014,furst_microrheology_2017}. The technique of 'microrheology' provides insight into local material properties and thus extends macrorheological investigations. Recently, such investigations tracking colloidal probe particles were performed in model complex fluids, where the macroscopic rheological response is rather well characterized. An example are  wormlike micellar solutions, for which local flow curves, i.e.~nonlinear force-velocity relations~\cite{gomez-solano_probing_2014,jain2021two}, 
particle oscillations during shearing~\cite{Berner2018,jain2021two,jayaraman_oscillations_2003,handzy_oscillatory_2004}
and transient particle motion \cite{Gomez-Solano2015-qu,Ginot2022recoil} were investigated. The existence of a number of different relaxation channels for the probe motion was recorded, which could be considered an intrinsic property of the system of viscoelastic fluid plus immersed colloidal particle. In contrast, in dense colloidal suspensions \cite{wilson_microrheology_2011,harrer_force-induced_2012,gazuz_active_2009,Squires2005}, a delocalization transition at finite forcing strength was discovered \cite{senbil_observation_2019}. At a critical force, the probe decouples from the surroundings and its motion records very atypical bath particle fluctuations. 

The friction force experienced by an individual Brownian particle and the velocity by which it moves relative to a Newtonian solvent are related by a friction coefficient $\gamma$. Alternatively, the velocity the particle attains when subject to a force can be written in terms of a mobility $\mu$. For the mentioned case of Newtonian solvent, mobility and friction coefficients are each other's inverses and are also connected to the diffusion coefficient via temperature in the famous Stokes-Einstein-Sutherland relation  $D_0=\kbt \mu=\frac{\kbt}{\gamma}$; here $k_B$ is Boltzmann's constant \cite{Dhont}. 

In viscoelastic fluids, where memory matters, it is well known that both coefficients generalize to time-dependent (retarded) kernels whose time-dependence encodes the temporal correlations of the fluid \cite{hansen2013theory,gotze2009complex}.
Rearrangements of the fluid take longer with e.g.~increasing concentration and thus forces at earlier times still influence the velocity at the present time. A similar memory of motion at earlier times also affects the friction force at present.
Regarding the Einstein relation, the kernels are then in general no longer related~\cite{Squires2005, saito_complementary_2017}, not even at zero frequency, i.e., large times. For example, the friction memory kernel may depend on the confinement of a probe particle~\cite{daldrop_external_2017,Kowalik2019,muller_properties_2020,basu_dynamics_2022}.
There is so far no complete understanding  how the time-dependencies of  mobility and friction kernels are related.

Recent microrheology experiments of colloidal probes in worm-like micellar solutions indicated that recoil spectra could provide crucial insights \cite{Gomez-Solano2015-qu,Ginot2022recoil}. It is known that such micellar solutions in the semi-dilute regime are well described by a Maxwell model with a single timescale \cite{Rehage1988,Spenley1993,Berret1997,cates1990statics}; at higher concentrations more complex relaxation behavior can occur~\cite{Jeon_2013,baiesi2021}. The Maxwell model captures the memory on macroscopic scales such as macrorheological measurements. Moreover, previous works could systematically fit the equilibrium Brownian motion, for example the mean squared displacement, of immersed colloids to Maxwell's model with a single relaxation time \cite{Ginot2022recoil,Gomez-Solano2015-qu,Ginot2022,van_zanten_brownian_2000,lu_probe_2002,vandergucht2003}. 
Yet, 'recoil' measurements, which test the back-motion of the colloidal probe when released after it was moved with optical tweezers, recorded a very different temporal evolution. Most notably, the recoil dynamics exhibit at least two timescales, with the faster one being much shorter than Maxwell's macroscopic relaxation time~\cite{Gomez-Solano2015-qu}. Experiments of partial loading and partial relaxation (close to equilibrium) showed the same two timescales, which generates the question how recoil and equilibrium mean squared displacement are related. 

In the present work, we build on these findings and study equilibrium mean squared displacements of a probe and its recoils after weak drivings. Applying concepts from linear response theory enables us to determine the relation between the two experiments. We will find that one (the equilibrium mean squared displacement) is dominated by the retarded friction kernel, whose integral grows strongly with density, while the other (recoils) provides direct access to the mobility kernel. Analysing two theoretical models, this explains the differing behaviors recorded in either experiments.

The manuscript is organized as follows: we start with the derivation of a linear response relation between recoil and MSD in section~\ref{chap:LinearResponse}.
In section~\ref{chap:ExperimentalSetup} we introduce the experimental system and in section~\ref{chap:ExperimentalComparison} compare experimental MSD and experimental recoil, i.e.~test the linear response relation.
The analysis of two theoretical models, a linear two-bath particle model and mode coupling theory, is presented in section~\ref{chap:ViscoelasticModels}.
In section~\ref{chap:Discussion} we discuss the relation between these two models and explain the observed concentration-dependence of recoil and MSD.


\section{Linear response relation between recoil and MSD}\label{chap:LinearResponse}

\begin{figure}
    \centering
    \includegraphics{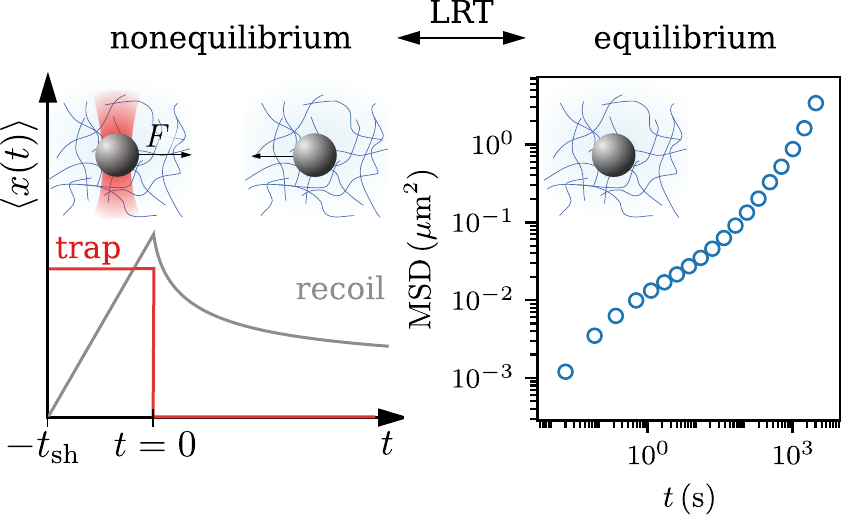}
    \caption{Left: Sketch of transient recoil experiment (gray line) after a colloidal probe particle in a complex fluid was perturbed with force $F$ (red line), applied  e.g. with optical tweezers (inset), for a time $t_\sh$. Right: Equilibrium mean squared displacement (MSD) of a colloidal particle in a wormlike micellar solution (CPyCl/NaSal \SI{7}{\milli \Molar}, open blue symbols). The two experiments are related via a linear response relation. }
    \label{fig:F1}
\end{figure}

In this work we theoretically and experimentally study the motion of a colloidal probe particle in a viscoelastic fluid, namely a wormlike micellar solution, keeping in mind that the theoretical analysis is valid for other viscoelastic fluids, such as dense colloidal suspensions, as well. Specifically, we are interested in the experimental setups depicted in Fig.~\ref{fig:F1}. Its left panel shows so called recoils, where the particle is dragged by a force  for a certain amount of time, and is then released at time $t=0$. Due to the deformed microstructure and accumulated strain  in the viscoelastic fluid, after releasing, the particle performs a backward motion opposite to the direction of the forced motion. We define the (positive) recoil $\recoil$ as
\begin{align}
    \recoil \equiv - \left[ \av{x(t)} - \av{x(0)}  \right],
    \label{eq:recoil_definition}
\end{align}
with $\langle\dots\rangle$ denoting averaged quantities.
The right hand side of  Fig.~\ref{fig:F1} shows the mean squared displacement $\MSD (t) \equiv \av{[x(t)-x(0)]^2}_\eq$~\cite{hansen2013theory}, obtained in absence of any driving.

Here, we derive a relation between these quantities. This derivation is explicit, based on a microscopic system of identical spherical particles, obeying overdamped dynamics~\cite{Dhont}. However, the final equation, Eq.~\eqref{eq:RecoilMSDF}, encodes the general form of the fluctuation-dissipation-theorem (FDT)~\cite{hansen2013theory}, and is thus valid in more general settings.

One of the particles is considered as probe particle. The free diffusion coefficient is $D_0=k_B T/\gamma$ and the particles interact via forces $\Fi = - \del_i V$ where quantities without index, such as $\mathbf{F} = - \del V$, refer to the probe particle, and $\del$ is a spatial gradient. 
Perturbing the probe with a small time-dependent force $\Fex(t)$, leads to the Smoluchowski operator \cite{Dhont}
\begin{equation}
	\Omega(t) = \Omega_\eq + \frac{D_0}{\kbt}\Fex(t) \cdot (-\del)\;.
\end{equation}
The solution for the probability density is split in a similar fashion
\begin{equation}
	\Psi(t) = \Psi_\eq + \delta\Psi(t)\;, 
\end{equation}
where $\Psi_\eq \propto \exp(-V/\kbt)$ and $\delta \Psi$ arises from $\Fex$.
This results in the linearized Smoluchowski equation for the deviation of the probability distribution
\begin{equation}
	\partial_t \delta\Psi(t) = \Omega_\eq \delta\Psi(t) - \frac{D_0}{\kbt}\Fex(t)\cdot\del\Psi_\eq + \mathcal{O}(F_\ex^2)\;.
\end{equation}
The solution is (with $\beta = 1/\kbt$)
\begin{equation}
	\delta\Psi(t) = - \frac{\beta}{\gamma} \int_{-\infty}^t dt' \Fex(t') \cdot e^{(t-t')\Omega_\eq} \mathbf{F} \Psi_\eq\;.
\end{equation}
Using this to calculate the linear response of an observable $A$ to the external force we obtain a form, which may be familiar from the fluctuation dissipation theorem,
\begin{equation}
	\av{A(t)}^{\lr}  = \av{A}_\eq - \frac{\beta}{\gamma} \int_{-\infty}^t dt' \Fex(t') \cdot \av{ \mathbf{F} e^{(t-t')\Omega_\eq^\dagger} A}_\eq.
\end{equation}	
In this isotropic system, we assume, without loss of generality, the external force to point in $x$-direction.
We are interested in the displacement of the particle $\av{x(t)}^{\lr}$, linear in applied force. For symmetry reasons, only its $x$ component is finite. Taking its time derivative results in the application of the Hermitian conjugate $\Omega^\dagger$ of the Smoluchowski operator on the variable $x$. We obtain
\begin{align}
   \partial_t \av{x(t)}^{\lr} = \gamma^{-1} F_\ex(t) - \int_{-\infty}^t dt' \gamma^{-1} F_\ex(t')M(t-t'),
    \label{eq:MobilityKernelEq}
\end{align}
where
\begin{equation}
    M(t-t') = D_0 \av{\beta F_x e^{(t-t')\Omega_\eq^\dagger} \beta F_x}_\eq\label{eq:M}
\end{equation}
is the \emph{mobility-kernel}, the correlation of interaction force felt by the probe particle, evaluated in equilibrium. 
	
The desired connection to the equilibrium mean squared displacement (MSD) starts from~ \cite{NAGELE1996215, hansen2013theory} 
\begin{equation}
	\partial_t \MSD(t) = 2D_0 - 2D_0\int_0^t ds' M(s').
	\label{eq:MMSD}
\end{equation}
Comparing Eq.~\eqref{eq:MMSD} with \eqref{eq:MobilityKernelEq} lets us conclude
\begin{equation}
	\av{x(t)}^\lr = \frac{1}{2\kbt} \int_{-\infty}^t dt' \partial_t\MSD(t-t') F_\ex(t')
	\label{eq:LinearResponse}
\end{equation}
which  can be checked by reinserting. This result is the central linear response relation generalizing Einstein's relation between friction and diffusion coefficients to arbitrary time dependencies of the mean drift and the quiescent mean squared diffusion. 

Upon inserting this into the definition of a recoil, Eq.~\eqref{eq:recoil_definition}, we obtain for the recoil in linear response
 \begin{equation}
 \begin{split}
      \recoil  &= \frac{1}{2 k_B T} \int_{-\infty}^0 dt' F_\ex(t') \partial_{t'} \left[\MSD(t-t')-\MSD(-t') \right]  \\
      \quad &- \frac{1}{2 k_B T} \int_{0}^{t}dt' F_\ex(t') \partial_t \MSD(t-t') .
      \end{split}
     \label{eq:RecoilMSDF(t)}
 \end{equation}
In the case of a time independent  force acting from $t=-t_\sh$ to $t=0$, as used in our experiments,
the recoil simplifies to ($t>0$)
\begin{align}
    \recoil = \frac{ F_\ex}{2 k_B T} \left[ \MSD(t) + \MSD(t_\sh) - \MSD(t+t_\sh) \right].
\end{align}
For ergodic fluids, where the MSD is diffusive for large correlation times, we define the 
long-time diffusion coefficient $D_\infty$ as
\begin{align}
      \lim_{t\to \infty} \frac{d}{dt} \MSD(t) =: 2 D_\infty .
    \end{align}
With this, the recoil takes an even simpler relation in the limit of long shear times, $t_\sh\to\infty$ (corresponding to the limit where the probe attained its stationary drift velocity before release),
\begin{align}
    \recoil = \frac{F_\ex}{2 k_B T} \left[ \MSD(t) - 2 D_\infty t \right].
    \label{eq:RecoilMSDF}
\end{align}
Eq.~\eqref{eq:RecoilMSDF} shows a relation between recoil and MSD, and interestingly it is the MSD minus its long time asymptote that appears.  Eq.~\eqref{eq:RecoilMSDF} also already hints  that recoil experiments, probing an interesting non-equilibrium aspect of the system, may give important insights into the equilibrium dynamics as well.


\section{Test in micellar solutions}\label{chap:MicellarTest}

Having established a relation between MSD and the recoil, we aim to apply it to experimental data obtained in wormlike micellar solutions.

We note right away that our experiments use optical tweezers, and are not performed by controlling the driving force, so that the requirements for  Eq.~\eqref{eq:RecoilMSDF} to be valid are not strictly given.

\subsection{Experimental setup}\label{chap:ExperimentalSetup}

In the experiments we analyzed viscoelastic equimolar cetylpyridinium chloride monohydrate (CPyCl) and sodium salicylate (NaSal) solutions with concentrations ranging from $\SI{5}{\milli \Molar}$ to $\SI{9}{\milli \Molar}$ (recoil measurements are limited to $\SI{7}{\milli \Molar}$),
which are contained in a $\SI{100}{\micro \m}$ thick sealed sample cell.
The cell was kept at a constant temperature of $\SI{25}{\celsius}$, leading to the formation of an entangled network of giant worm-like micelles~\cite{cates1990statics}. 
To probe the fluid's microrheology, we suspended a small amount of silica probe particles with diameters $\SI{2.73}{\micro \m}$ in the solution. 

For the measurement of recoils the colloidal probe is trapped in an optical tweezer, built of a Gaussian laser ($\lambda = \SI{1064}{\nano \m}$) and a high magnification microscope objective ($100\times$, $\mathrm{NA} = \SI{1.45}{}$).
Using relatively large trapping strengths (see Table~\ref{tab:ExpParam} in Appendix~\ref{chap:ExperimentalParameters}), the probe is positioned in the center of the trap and we can apply a constant-velocity perturbation via a relative motion between the probe/trap and the fluid.
This is achieved by a computer controlled piezo-driven stage that translates the sample cell with constant velocity $v$ and that is synchronized with the laser intensity.
The time of translation or shear time $t_\sh$ is chosen sufficiently long, such that the probe reached a non-equilibrium steady state before release.
To avoid interactions with the sample walls, the trap was located at least $\SI{30}{\micro \m}$ away from any surface.
We extract the probe's trajectory by video microscopy with a frame rate of $\SI{100}{\Hz}$, which, using a custom Matlab algorithm~\cite{crocker1996methods} yields an accuracy of $\pm \SI{6}{\nano \m}$ of the particle position.
For further details regarding the experimental recoil setup, we refer to the recent work of Ginot~\textit{et.~al}~\cite{Ginot2022recoil} and its Supplementary Material.

To test the linear response relation, Eq.~\eqref{eq:RecoilMSDF}, we also measured the mean squared displacement of freely diffusing colloidal particles suspended in the micellar solution, and at the surface of the sample cell. The 2-dimensional trajectories $\mathbf{r}(t) = (x(t),y(t))^T$ of the colloidal particles were extracted using video microscopy and above mentioned Matlab algorithm.
Due to the system's isotropy, we calculated the (one-dimensional) MSD, after subtracting drift from the trajectories, and by averaging over both dimensions,
\begin{align}
    \MSD(t) \equiv \frac{1}{2} \av{\left|\mathbf{r}(t_0+t)-\mathbf{r}(t_0)\right|^2}_{t_0}.
    \label{eq:2DMSDExp}
\end{align}
Finally, we run an ensemble average over several particles ($\sim 20$) for better statistics.
We checked that presence of the glass surface has negligible influence on the reported MSD, by comparing with bulk measurements.

\subsection{Comparison of experimental MSD and experimental recoil via Eq.~\eqref{eq:RecoilMSDF}}\label{chap:ExperimentalComparison}

\begin{figure}
    \centering
   \includegraphics[]{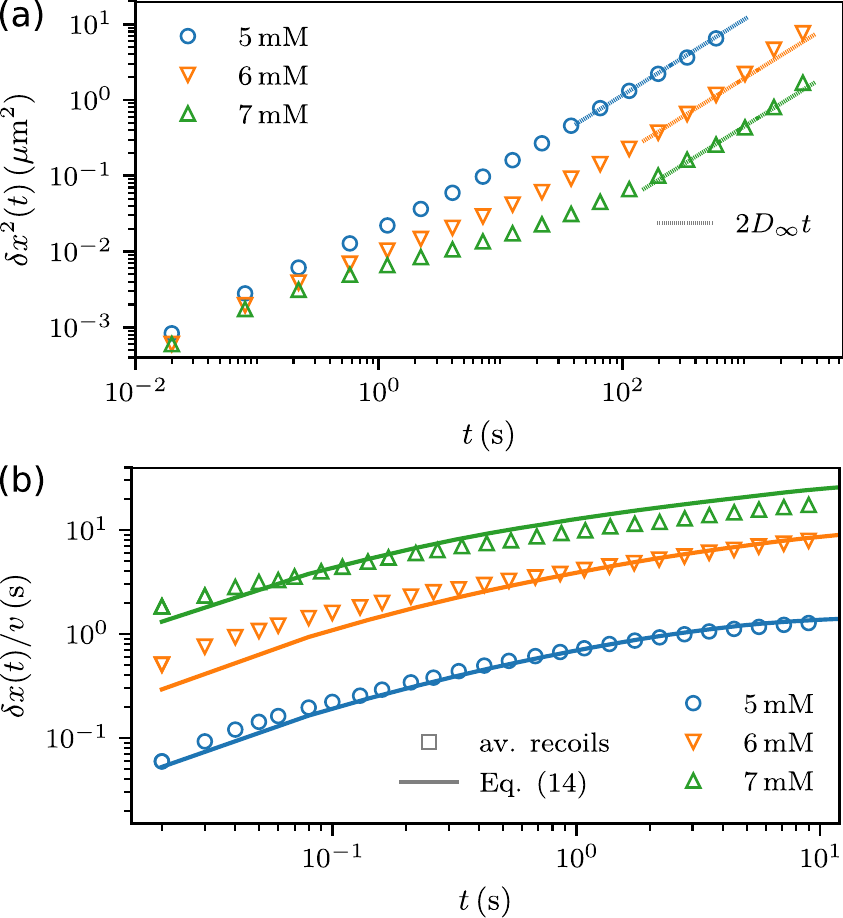}
    \caption{
       (a) Experimental mean squared displacements (open symbols) measured in wormlike micellar solutions with concentrations of 5 (blue), 6 (orange) and $\SI{7}{\milli \Molar}$ (green). The colored dotted lines highlight the long-time diffusion $2 D_\infty t$. (b)  Experimental test of Eq.~\eqref{eq:RecoilMSDVNESS}: experimental mean recoils averaged over different shear velocities in the linear regime (left-hand side of Eq.~\eqref{eq:RecoilMSDVNESS}, open symbols) nicely match recoil curves calculated from the MSDs in (a), i.e.~right-hand side of Eq.~\eqref{eq:RecoilMSDVNESS} (solid lines).
    Note that the $x$-axes in (a) and (b) have different scaling as recoils saturate quickly and the measurement was limited to a time interval of $\SI{9}{\s}$. 
      }
    \label{fig:TestRecoilRelation}
\end{figure}

In Fig.~\ref{fig:TestRecoilRelation}(a) we show experimental MSDs for concentrations of 5, 6 and $\SI{7}{\milli \Molar}$, together with asymptotic fits of the long-time diffusion, $\lim_{t\to \infty}\MSD(t) = 2 D_\infty t$ (dotted lines).
Clearly, with increasing concentration the sub-diffusive plateau at intermediate times becomes strongly pronounced, while the long-time diffusion coefficient is reduced by almost two orders of magnitude from 5 to  $\SI{7}{\milli \Molar}$.
In contrast, at very short timescales all curves nearly collapse, as we expect the colloidal probe to diffuse with a short-time diffusion coefficient that is almost independent of micellar density \cite{Dhont}. 

Before comparing experimental MSDs and recoils we come back to the above note, and to the fact that the linear response relation, Eq.~\eqref{eq:RecoilMSDF}, describes recoils after a constant-force perturbation. 
In contrast, the optical trapping strength of our experimental recoil system is very high, such that the probe particle almost completely follows the shear velocity $v$ of the trap.
This perturbation by constant velocity, however, is generally not equivalent to a constant-force perturbation~\cite{Squires2005}.
Specifically, for our viscoelastic systems the friction memory kernel reveals a trap stiffness dependence~\cite{muller_properties_2020}, meaning that the weak (constant-force) and strong (constant-velocity) trap limits are generally not comparable. 
This is why we restrict this comparison to the three lowest micellar densities of our studies, as the mentioned differences between driving modes are expected to be the more pronounced the higher the density. We thus employ the Einstein relation $v= \frac{D_\infty}{\kbt} F_\ex$ to replace the external force in Eq.~\eqref{eq:RecoilMSDF} with the stationary velocity $v$, resulting in
\begin{equation}
    \frac{\recoil}{v} = \frac{\MSD(t) - 2 D_\infty t}{2 D_\infty}.
    \label{eq:RecoilMSDVNESS}
\end{equation}
The resulting test of Eq.~\eqref{eq:RecoilMSDVNESS} is shown in Fig.~\ref{fig:TestRecoilRelation}(b).
Open symbols show experimental recoils normalized by shear velocity $v$.
For better statistics, these curves are averages of several recoils measured at different shear velocities $v$, within the linear regime where $\av{x(t)}\propto v$.
The solid lines in Fig.~\ref{fig:TestRecoilRelation}(b) show the result of applying Eq.~\eqref{eq:RecoilMSDVNESS} to the experimental MSDs shown in Fig.~\ref{fig:TestRecoilRelation}(a).
Clearly, for all concentrations, recoils and the result of applying Eq.~\eqref{eq:RecoilMSDVNESS} to experimental MSDs agree reasonably well, thereby confirming Eq.~\eqref{eq:RecoilMSDVNESS}.
Deviations at very short times $t \lesssim \SI{0.1}{\s}$ are related to an imperfect trigger of the beginning of recoils.

This corroborates that recoil and MSD are
directly related in linear response. They are characterized by the same set of timescales and corresponding relative amplitudes.
Notably compared to the strong {\it qualitative} density dependence of the MSD, one can hardly differentiate recoils at different concentrations. The recoils shown in Fig.~\ref{fig:TestRecoilRelation}(b) can almost be brought on top of each other by an overall factor, i.e., the recoil curves hardly change qualitatively with density.
We will discuss this observation in detail in the next sections.


\section{Analysis in models of viscoelastic fluids} \label{chap:ViscoelasticModels}

In this section, we discuss two theoretical models that yield good fits of experimental MSDs and recoils at increasing concentrations up to $\SI{9}{\milli \Molar}$. 
These models have complementary benefits. We start with  a linear two-bath particle model which has been previously proposed for a quantitative understanding of recoils~\cite{Ginot2022recoil}. It is very generic and can be analyzed in complete analytical detail~\cite{doerries2021correlation}. 
We then turn to mode coupling theory (MCT) of monodisperse hard spheres. It is based on a microscopic starting point, and extends to the nonlinear regime. Its most detailed predictions concern the  glass transition which lies beyond the experimental regime; this discussion thus is relegated to an Appendix.

\subsection{Two-bath particle model}\label{chap:2BP}

\begin{figure}
    \centering
    \includegraphics[]{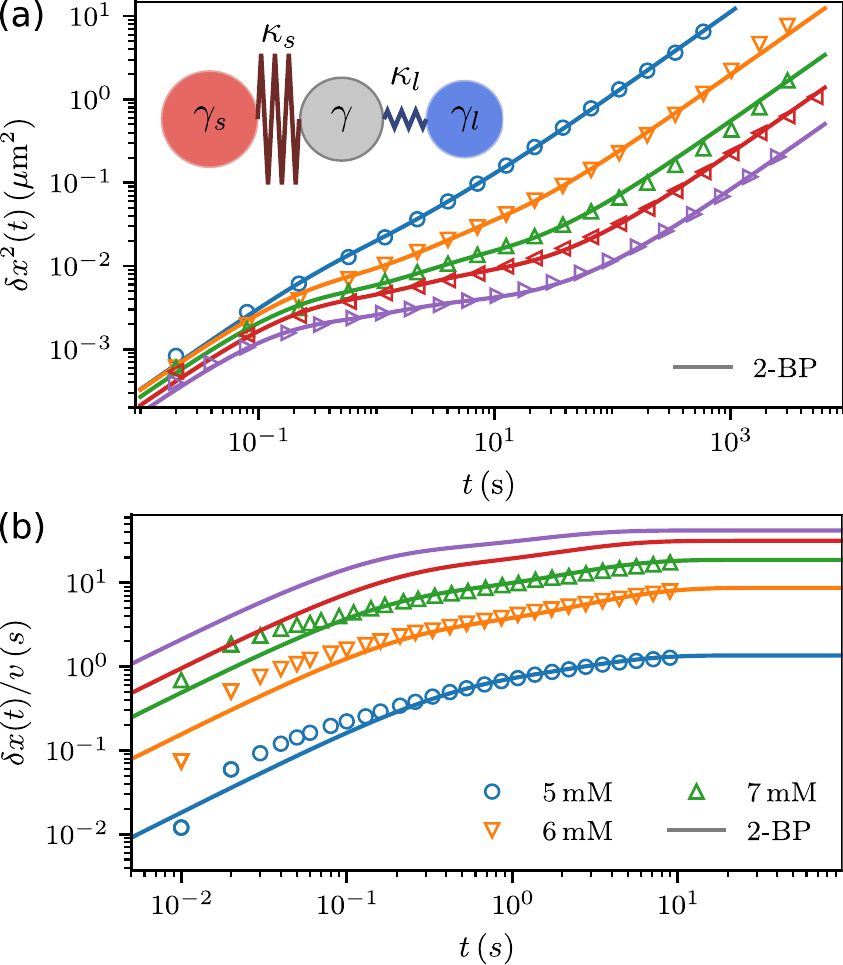}
    \caption{Mean squared displacement and recoils within a linear two-bath particle (2-BP) model and from experiments. 
    (a) Inset: Sketch of the linear two-bath particle model. A probe particle with friction coefficient $\gamma$ is coupled via harmonic springs of stiffness $\kappa_s$ and $\kappa_l$ to two bath particles with friction coefficients $\kappa_s$ and $\kappa_l$, respectively.
    Main graph: Experimental MSDs (colored open symbols) for concentrations between $\SI{5}{\milli \Molar}$ (blue) and $\SI{9}{\milli \Molar}$ (purple) together with linear two-bath particle models (lines, see Tab.~\ref{tab:MSD2BPFit} for model parameters). 
    (b) Experimental recoils (colored open symbols) for concentrations ranging from $\SI{5}{\milli \Molar}$ (blue) to $\SI{7}{\milli \Molar}$ (green) together with respective 2-BP models. The red and purple line correspond to 2-BP recoils matching experimental MSDs at 8 and $\SI{9}{\milli \Molar}$ in (a).
    Note that the time axes are scaled differently in both plots.}
    \label{fig:2BP}
\end{figure}

The experimental recoils in Fig.~\ref{fig:TestRecoilRelation}(b) reveal several relaxation processes (the curves allow to distinguish at least two). 
Indeed, previous work reported double-exponential behavior with two distinct timescales~\cite{Gomez-Solano2015-qu,Ginot2022recoil}. 
As our experiments concern the linear regime, where recoil amplitudes are linear in shear velocity $v$, the simplest model that generates the observed behavior is a linear two-bath particle model, as sketched in the inset of Fig.~\ref{fig:2BP}(a). 
It consists of a probe particle with friction coefficient $\gamma$ that is coupled to two bath particles with friction coefficients $\gamma_s$ and $\gamma_l$ via harmonic springs of stiffness $\kappa_s$ and $\kappa_l$, respectively.
This microscopic model can be rationalized as an extension of the Maxwell model, the latter describing a single timescale of macroscopic processes.
While the Maxwell model is known to describe the equilibrium properties of wormlike micellar systems very well~\cite{Spenley1993,Rehage1988,Berret1997,cates1990statics}, the situation is different when a probe particle is sheared through the fluid.
Then the probe couples to the fluid, thereby apparently  inducing a second timescale, which is  modeled via a second bath particle.
We proceed with a brief introduction of the main findings of Ginot \textit{et. al}~\cite{Ginot2022recoil}.

The dynamics of the probe (recall that probe variables carry no subscript) and two bath particles is described via a set of overdamped Langevin equations ($i=s,l$) 
\begin{align}
\begin{split}
    \gamma \dot{x}(t) &= F_\trap(t) - \sum_{i} \kappa_i \left[x(t)-x_i(t)\right]+\xi(t),
\end{split} \label{eq:Langevin_tracer}
    \\
    \gamma_i \dot{x}_i(t) &= \kappa_i \left[ x(t) - x_i(t) \right] + \xi_i(t), \label{eq:Langevin_bath}
\end{align}
where the probe particle experiences a trapping force $F_\trap(t)=-\kappa(t) \left[ x(t) - x_0(t)\right]$ due to the optical potential of stiffness $\kappa(t)$ and position $x_0(t) = \int^t dt' v(t')$~\cite{Ginot2022recoil}.
$\xi$ and $\xi_i$ denote independent Gaussian white noises, i.e. for $(\xi_i,\xi_j)\in \{\xi,\xi_s,\xi_l\}$~\cite{Ginot2022recoil}
\begin{equation}
    \langle \xi_i(t)\rangle = 0, \quad \langle \xi_i(t) \xi_j(t') \rangle = \delta_{ij} 2 k_\mathrm{B}T \gamma_i \delta(t-t'). \label{eq:Langevin_noise}
\end{equation}
Because the noise strength is set by a unique temperature $T$, the model encodes the fluctuation dissipation theorem. 

Characterizing the probe's trapping, there exist two extreme cases: $\kappa \to \infty$ and $\kappa = 0$, which are characterized by different sets of eigenmodes in the two-bath particle model \cite{Ginot2022recoil}.
If $\kappa \to \infty$, the probe is pinned by the trap, and its motion is deterministic.
The two bath particles are then de-coupled in the stochastic sense. Each bath particle interacts with the deterministic probe. The interaction between probe and bath particles is then termed  \textit{nonreciprocal}, as the probe's motion is not altered by the forces between probe and bath particles.
In this case the respective nonreciprocal timescales $\tilde\tau_{s,l}$ equal the individual relaxation times of the bath particles, i.e.~$\tilde\tau_i = \gamma_i/\kappa_i$ for $i=s,l$~\cite{Ginot2022recoil}.

In the other limit, for $\kappa=0$, the probe couples the two bath particles and force balance applies. The probe's motion is altered by the forces from bath particles, and the interaction is termed \textit{reciprocal}. The relaxation timescales contain all couplings and read 
($\zeta_i=(\gamma+\gamma_i)/\tilde\tau_i$)
\begin{align}
	\tau_{s,l}^{-1} =\frac{1}{2 \gamma }\left[\sum_i\zeta_i \pm 
\sqrt{(\sum_i\zeta_i)^2+4(\kappa_s\kappa_l-\zeta_s\zeta_l)}\right],\label{tauSLAnalytic}
\end{align}
where the positive sign corresponds to the shorter timescale $\tau_s$ and the negative sign to the longer timescale $\tau_l$~\cite{Ginot2022recoil}.

Naturally, both recoil and MSD are measured when the optical trap is off ($\kappa = 0$). Thus, they are expected to be characterized by the reciprocal timescales $\tau_{s,l}$.
Indeed, this is confirmed by the analytical expressions for the recoil $\recoil$ and the MSD $\MSD(t)$,
\begin{align}
\recoil &= A_\mathrm{tot}-A_s e^{-\frac{t}{\tau_\text{s}}}-A_l e^{-\frac{t}{\tau_\text{l}}} \quad \text{and}  \label{RecoilTheo} \\
\MSD (t) &= 2D_\infty t + A_\mathrm{tot}'-A_s'e^{-\frac{t}{\tau_s}} - A_l' e^{- \frac{t}{\tau_l}},
    \label{eq:MSD}
\end{align}
where $D_\infty = \frac{k_B T}{\gamma + \gamma_s+\gamma_l}$~\cite{Ginot2022recoil}.

The amplitudes in  Eq.~\eqref{RecoilTheo}, $A_{s,l,\mathrm{tot}}$, depend on shear time $t_\sh$ (see Appendix~\ref{chap:RecoilAmplitudes2BP} for full expressions). In the limit of infinite shear time ($t_\sh \to \infty$), i.e.~starting from a nonequilibrium steady state, the recoil amplitudes approach a constant value and we find the following relation between  amplitudes appearing in recoil and those of MSD, 
\begin{align}
     A'_{s,l,\mathrm{tot}}&=\lim_{t_\sh\to\infty}\frac{2 D_\infty A_{s,l,\mathrm{tot}}}{v}.\label{eq:equal}
\end{align}
Here, $v$ denotes the (constant) velocity with which the probe is sheared.
Moreover, the recoil amplitudes $A_{s,l,\mathrm{tot}}$ are linear in $v$ due to the linearity of the two-bath particle model.
Inserting the model's (linear) force-velocity relation $F_\ex = v \left(\gamma + \gamma_s+\gamma_l \right) = v \frac{k_B T}{D_\infty}$, we confirm the linear response relation~\eqref{eq:RecoilMSDF} between recoil and MSD within the 2-BP model. 
Indeed, for this model with {\it linear} couplings, a number of simplifications occur: The recoil, for $t_\sh\to\infty$, is independent of $\kappa$, i.e., recoils after force driving and velocity driving are identical. This is why Eq.~\eqref{eq:RecoilMSDF} applies in any driving mode.  

With the analytical expressions at hand, experimental data can be fitted. This is done for concentrations ranging from $\SI{5}{\milli \Molar}$ to $\SI{9}{\milli \Molar}$. For each density, a single set of parameters is obtained fitting recoil and MSD, see Tab.~\ref{tab:MSD2BPFit} in Appendix~\ref{chap:ModelParameters}, while parameters naturally depend on micellar density. 
The resulting MSD and recoil curves are shown in Figs.~\ref{fig:2BP}(a) and (b) as open symbols (experiments) and solid lines (2-BP model) and show remarkable agreement.
Note that for high concentrations of 8 and $\SI{9}{\milli \Molar}$, only experimental data of MSDs are shown (and fitted), and no data for recoils are presented. The reason is that these high densities pose several difficulties; The resulting high viscosities require large trap stiffness, and large laser powers may lead to local heating. Also, very small shear velocities are required to stay in the linear regime. Last, Eq.~\eqref{eq:RecoilMSDF} fails when applied to velocity controlled driving, for the reasons given above. Fitting MSDs is still insightful as it allows to see how the obtained timescales vary with density.
Also, we still applied Eq.~\eqref{eq:RecoilMSDVNESS} for these cases to be able to formally obtain the curves (red and purple lines) in Fig.~\ref{fig:2BP}(b). 

As an interesting remark when comparing Eqs.~\eqref{RecoilTheo} and \eqref{eq:MSD}, and using Eq.~\eqref{eq:equal}, it becomes notable that the MSD provides one additional fitting constraint compared to recoil, through the long-time diffusion coefficient $D_\infty$. Indeed, fitting recoils alone leaves one parameter of the model undetermined, e.g., the probe's friction $\gamma$. This has a physical reason: In reciprocal motion, only relative forces and relative motion is important, i.e., the center of mass motion of the three particles is not detected. In long-time diffusion this center of mass motion is however detected, and thus one more constraint is provided by the MSD.    

\begin{figure}
    \centering
    \includegraphics{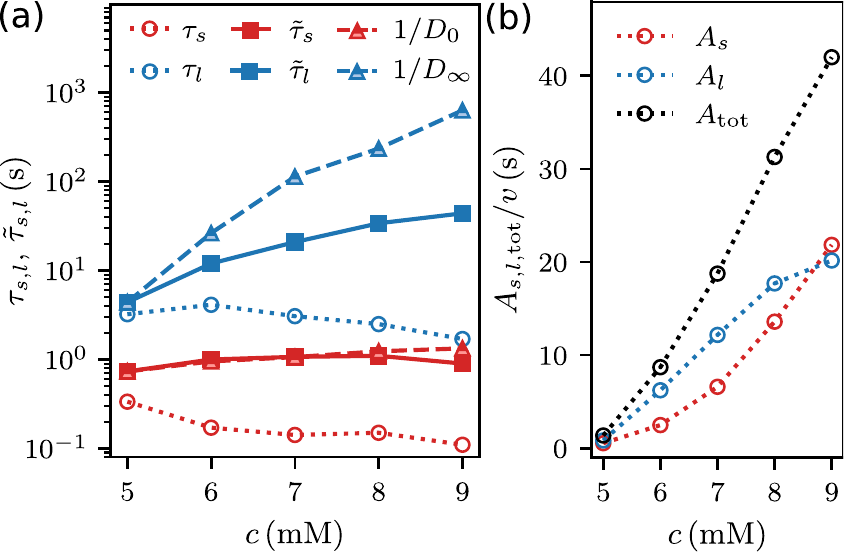}
    \caption{(a) Timescales $\tau_{s,l}$ (open circles) and $\tilde\tau_{s,l}$ (solid squares) obtained from fitting experimental MSDs and recoils to two-bath particle models for various concentrations $c$. Additionally, we show the experimental short- and long-time inverse diffusion coefficients $1/D_{0,\infty}$ (half-filled triangles; see fits in Fig.~\ref{fig:TestRecoilRelation}(a). For a qualitative comparison they are given in arbitrary units and rescaled to match $\tilde\tau_{s,l}$ at $c = \SI{5}{\milli \Molar}$. (b) Corresponding recoil amplitudes $A_{s,l,\mathrm{tot}}$ normalized by shear velocity $v$.}
    \label{fig:TimeScales_RecoilAmplitudes}
\end{figure}

With the established agreement between experiment and model we proceed with an analysis of the model's timescales and amplitudes under variation of micellar concentration.
In Fig.~\ref{fig:TimeScales_RecoilAmplitudes}(a) we show reciprocal (open circles) and nonreciprocal (solid squares) timescales, $\tau_{s,l}$ and $\tilde\tau_{s,l}$, as a function of concentration $c$. 
Notably, the behavior of reciprocal and nonreciprocal timescales differs qualitatively.
While the reciprocal timescales remain almost constant or even slightly decrease upon increasing concentration, the larger nonreciprocal timescale $\tilde\tau_l$ increases by one order of magnitude.
Hence, depending on whether the probe is confined or free, the relaxation times either increase with density or remain almost constant.

This behavior has an intuitive reason, partly similar to the discussion of fitting constraints above; If the probe is strongly trapped or pinned (nonreciprocal case), the bath particles relax towards the probe. As their friction coefficients strongly increase with concentration (see Tab.~\ref{tab:MSD2BPFit}), the relaxation slows down.
In contrast, in the reciprocal case of a free probe, probe and bath particles relax  towards each other. This relative motion remains fast, as the probe's friction coefficient does not strongly grow. This interpretation carries over to the real experimental situation: When trapping the particle at a fixed position, the surrounding fluid needs to relax accordingly. This process is expected to be slower than the case when the probe particle can move freely, thereby possibly finding a faster path to equilibrium.  

How does the observed strong increase of the timescale $\tilde\tau_l$ compare to the significant reduction of the long-time diffusion coefficient $D_\infty$ observed in Fig.~\ref{fig:TestRecoilRelation}(a)?
Fig.~\ref{fig:TimeScales_RecoilAmplitudes} also shows  $1/D_\infty$, in arbitrary units, to map $\tilde\tau_{s,l}$ at $c=\SI{5}{\milli \Molar}$. This yields an additional timescale, i.e., the time the probe needs to diffuse a certain distance (e.g. its diameter). This timescale grows even stronger than the relaxation timescale of the nonreciprocal case,  approximately another order of magnitude compared to $\tilde\tau_l$! 
This is due to the fact that in $\tilde\tau_l$, both, the friction coefficient as well as the coupling strength, increase with concentration (see Tab.~\ref{tab:MSD2BPFit}), while $D_\infty$ is only anti-proportional to the zero-shear friction, i.e.~the sum of friction coefficients. Hence, $D_\infty$ exhibits a greater sensitivity with respect to changes in concentration. This interesting observation will be picked up again when discussing friction and mobility kernels below.
The also shown behavior of $1/D_0$ reveals (again) that the short time dynamics stays almost constant upon variations in concentration. 

Finally, in Fig.~\ref{fig:TimeScales_RecoilAmplitudes}(b) we show the recoil amplitudes $A_{s,l,\mathrm{tot}}$, compare Eq.~\eqref{RecoilTheo} and Appendix~\ref{chap:RecoilAmplitudes2BP}. 
In contrast to recoil timescales, the  amplitude $A_\mathrm{tot}$ increases by more than one order of magnitude for the density range shown, when normalized by shear velocity. This is partly due to the fact that, at fixed driving velocity, the driving force increases with density. 


\subsection{Mode Coupling Theory (MCT)}\label{chap:MCT}

Before specifying the second model provided by MCT, we address the relation between driving at given force to driving at given velocity.
In our experiments, the probe is not perturbed by force (basis of Eq.~\eqref{eq:RecoilMSDF}), but trapped by an optical tweezer that moves with constant velocity $v$, and we are looking for the way of replacing the force factor in Eq.~(\ref{eq:RecoilMSDF}) by this velocity. While a rigorous relation requires to map the dynamics including fluctuations, we proceed with an approximation that maps the average variables in the stationary state. As noted below Eq.~\eqref{eq:equal}, this is expected to become exact for systems with linear interactions, as fluctuations and averages then decouple.

Mathematically, the mapping is  achieved by introducing an irreducible operator in the Hermitian conjugate of the Smoluchowski operator $\Omega^\dagger$ \cite{CichockiHess,Gazuz2013}. Thereby, equation~(\ref{eq:MobilityKernelEq}) can be transformed into
\begin{equation}
	\gamma^{-1} F_\text{ex}(t) = v(t) + \int_{-\infty}^t dt' m(t-t')v(t')\;,
	\label{eq:FrictionKernelEq}
\end{equation}
where the equilibrium \emph{friction-kernel} $m$ is related to the mobility-kernel $M$ via a Volterra relation~\cite{tricomi_integral_1985}
\begin{align}
    \hat{M}(s) = \frac{\hat{m}(s)}{1+\hat{m}(s)}.
    \label{eq:Volterra}
\end{align}
$\hat{h}(s) = \int_0^\infty dt \, h(t) e^{-st}$ denotes the Laplace transform.
By definition, the friction-kernel contains irreducible dynamics and in the present case can not be reduced to a Green-Kubo form accessible to e.g.~simulation. 

Inserting the force that leads to a stationary average probe velocity $v$ in Eq.~\eqref{eq:FrictionKernelEq} into the linear response equation (\ref{eq:LinearResponse}) and considering the limit $t_\sh\to\infty$ leads to the recoil
\begin{equation}
    \frac{\recoil}{v} = \frac{\MSD(t) - 2 D_\infty t}{2 D_\infty}\;.
    \tag{\ref{eq:RecoilMSDVNESS}}
\end{equation}
This is the relation explicitly tested by experiments in Fig.~\ref{fig:TestRecoilRelation} which will be analyzed by MCT in the following.

\begin{figure}
    \centering
    \includegraphics{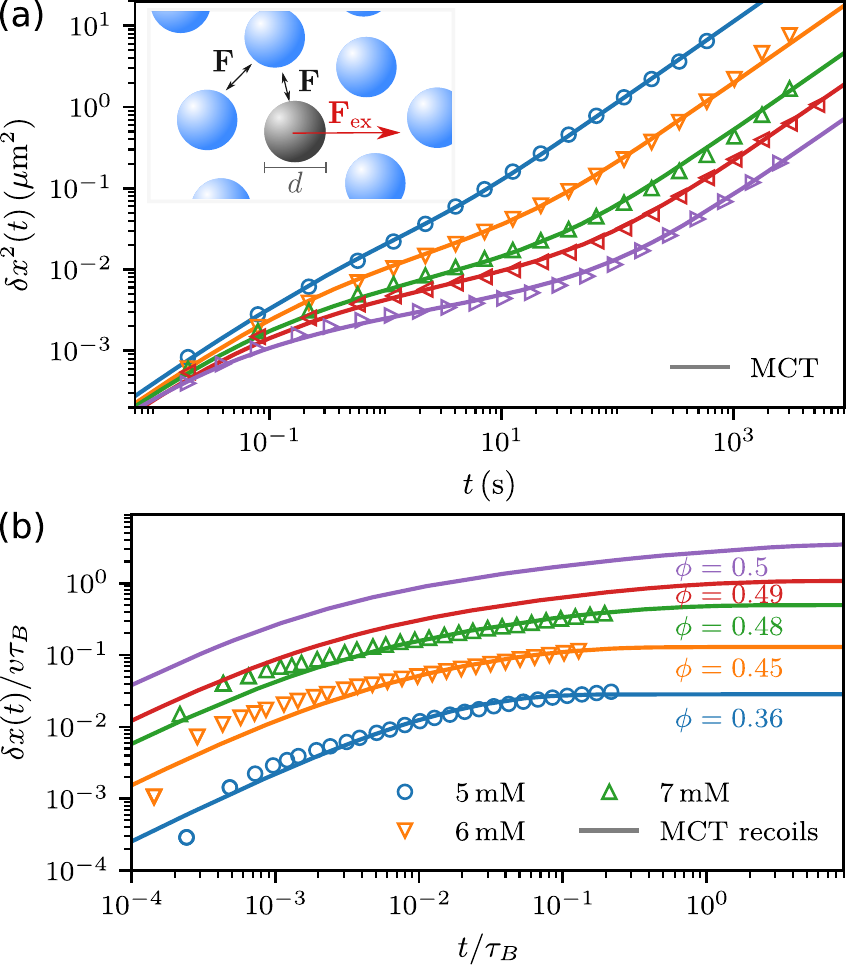}
    \caption{Mean squared displacement and recoils within MCT and from experiments. (a) Inset: Sketch of a system of monodisperse hard spheres of diameter $d$ that interact via interaction forces $\bf{F}$. The gray particle is considered as probe particle and perturbed with an external force $\mathbf{F}_\mathrm{ex}$. Main graph: Experimental MSDs (open symbols) fitted with MCT MSD's (lines) for different packing fractions $\phi$. The respective packing fractions $\phi$ are shown in (b) and were chosen such that MCT recoils (solid lines) match existing experimental recoils (open symbols, see also Fig.~\ref{fig:TestRecoilRelation}(b)). Note that the windows of time are different in the two plots as $\tau_B \approx \SI{50}{\s}$.} 
    \label{fig:MatchedMCT}
\end{figure}

As announced, as a second theoretical approach to the linear response of dragged colloidal probes,  we consider the MCT~\cite{gotze2009complex} of monodisperse hard spheres (see sketch in Inset of Fig.~\ref{fig:MatchedMCT}(a)). 
In MCT it is well established that the mean squared displacement fulfills the equation of motion \cite{fuchs1998asymptotic} (projected to one dimension)
\begin{equation}
    \MSD(t) + \int_0^t m(t-t') \MSD(t') dt' = 2 D_0 t
    \label{eq:MSD_m}
\end{equation}
with the friction-kernel $m(t)$ of Eq.~\eqref{eq:FrictionKernelEq} reappearing. In mode coupling approximation it is given as a functional of the probe and bath density correlators. The retardation of the friction kernel expresses that the friction the probe experiences is affected by the motion of the probe itself at earlier times and by the dynamics of the surroundings again at earlier times. Viscoelasticity is incorporated by the friction to develop longer-lasting memory with increasing concentration of the glass-forming solution. 
The system of coupled integro-differential equations resulting in MCT is solved by a preexisting MCT-solver \cite{Voigtmann2011} on a grid of 200 equidistant wavevectors $0.1,...,39.9$ using a decimation scheme with initial time step $10^{-8}$ and 128 steps per window \cite{Fuchs1991}. For the specified packing fraction the structure factor needed for $m(t)$ is chosen to be given by the Percus-Yevick approximation \cite{hansen2013theory}. The numerical MSDs from MCT are dimensionless of the form ${\MSD}^* (t^*)$ and can be shaped to a real MSD by $\MSD(t)=d^2 {\MSD}^* (t/\tau_B)$, where $d$ is the hard sphere diameter and $\tau_B = d^2/D_0$ denotes the Brownian time, which is the typical time a particle needs to diffuse its own diameter $d$~\cite{senbil_observation_2019}. Applying the recoil formula to such a curve gives ($2D_\infty^* \equiv \lim_{t\to\infty} \partial_{t^*} {\MSD}^* $)
\begin{equation}
    \frac{\recoil}{v} = \frac{{\MSD}^*(t^*)- 2D_\infty^* t^*}{2D_\infty^*} \tau_B     = X^*(t^*;\phi) \tau_B\label{eq:resc}.
\end{equation}
This means that plotting $\frac{\recoil}{v\tau_B}$ versus $t^*=t/\tau_B$ yields the function $X^*$, which for Brownian hard spheres only depends on the chosen packing fraction. The $D_\infty^*$ is obtained from a linear fit to the numerical curves in the last window of the decimation grid.

To match this theory to our experiments for a given micellar concentration, we start by determining the ratio $D_\infty/D_0$, which uniquely yields the corresponding hard sphere packing fraction $\phi$. 
We then fit the resulting MCT curves for recoils to experimental data, which determines  the Brownian timescale $\tau_B$ such that the scaled recoil lies on $X^*(t^*)$ defined in Eq.~\eqref{eq:resc}. Results are shown in Fig.~\ref{fig:MatchedMCT}(b). 
After this it is still possible to adjust the particle diameter to  match the MSD. As already observed below Eq.~\eqref{eq:equal}, fitting the recoil does not determine all parameters of the model.  
This analysis shows that it is possible to find parameters of a hard sphere system that describe the probe in the wormlike micellar solution including the recoil formula with good agreement.

For the concentrations $\SI{8}{\milli \Molar}$ and $\SI{9}{\milli \Molar}$, where no recoil was measured directly, the fit parameters $d^2$ and $\tau_B$ are both chosen to match just the MSD. 
The MSDs obtained by this procedure compared with the experimental MSDs are seen in Fig. \ref{fig:MatchedMCT}(a) while the fitting parameters are given in Tab.~\ref{tab:MSDMCTFit} in Appendix~\ref{chap:ModelParameters}. 

\begin{figure}
    \centering
    \includegraphics[]{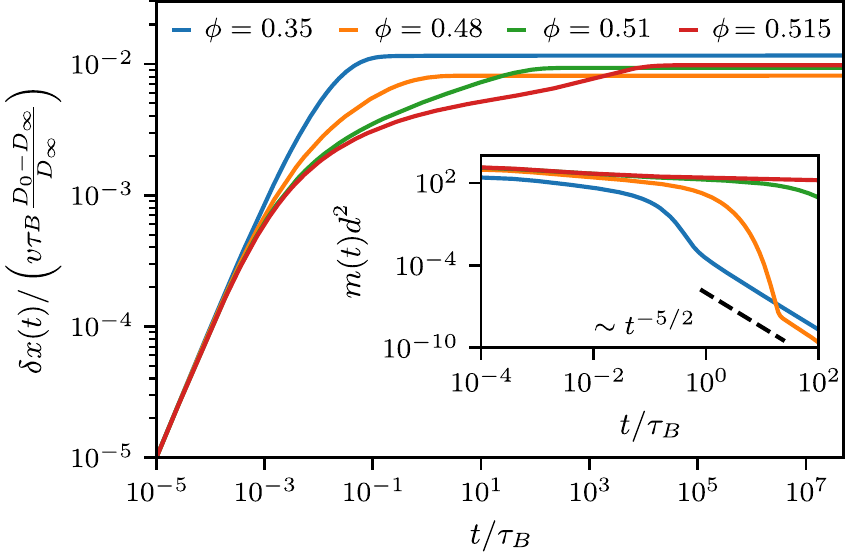}
    \caption{Recoils calculated from MCT MSD's via Eq.~\eqref{eq:RecoilMSDVNESS} and rescaled with respect to their short-time behavior. The first two packing fractions describe experimental measurements at $\SI{5}{\milli \Molar}$ and $\SI{7}{\milli \Molar}$ (up to $\sim 10^{-1}\tau_B$), respectively. For higher packing fractions, the systems approaches the glass transition. Inset: Corresponding friction-kernels $m(t)$ with long-time tails $\sim t^{-5/2}$~\cite{Mandal2019}.}
    \label{fig:ShortScaling}
\end{figure}

What other behavior of recoils can we extract? 
Fig.~\ref{fig:ShortScaling} shows theoretical recoil curves for a wide range of packing fractions and up to  very large values of $t^*= t/\tau_B$. They are re-scaled such that their linear short-time behavior overlaps, which is achieved by division by $(D_0-D_\infty)/D_\infty$. It can be seen that in the region $t^*\alt 10^{-1}$, identified as the region of our experimental measurements, the recoils have a very similar shape, especially for higher densities. This confirms our observation, that the timescales used in the two bath particle model hardly depend on density. 
At even larger times, the curves in Fig.~\ref{fig:ShortScaling} show an additional relaxation step, which moves to later and later times with increasing packing fraction. It can be analysed close to the glass transition of hard spheres for even higher packing fractions, presented in Appendix~\ref{chap:AlphaScaling}. For reference, the inset of Fig.~\ref{fig:ShortScaling} shows the retarded friction kernel. 
It exhibits a power-law long time tail with exponent $-5/2$  \cite{Mandal2019} which, however, is outside the experimental window where the recoils were determined.   

\section{Discussion}\label{chap:Discussion}
 
The observations on the relation between recoils and MSD can be formalized using memory kernels and thereby allow to draw a connection between the two theoretical models and to reveal universal aspects.

Linear response theory relates recoil to the mobility-kernel $M$. Integrating Eq.~\eqref{eq:MobilityKernelEq}    yields 
\begin{align}
   \recoil = \gamma^{-1} \int_0^t dt' \int_{-\infty}^{t'} ds' F_\mathrm{ex}(s') [\delta(t'-s')-M(t'-s')].\label{eq:recoil}
\end{align}
The mobility kernel measures the relaxation of forces, see Eq.~\eqref{eq:M}, with the probe particle moving (diffusing) freely: This is apparent in the derivation of   Eq.~\eqref{eq:M}. From our experiments, the corresponding timescales vary only weakly with density; we argued that the force free moving particle can relax fast. In the bath-particle model, this corresponds to the reciprocal case, and the mobility kernel $M$  contains the reciprocal timescales $\tau_{s,l}$, explicitly, 
\begin{align}
    M(t) = \frac{\gamma}{v (\gamma + \gamma_s+\gamma_l)} \left(\frac{1}{\tau_s^2} A_se^{-\frac{t}{\tau_s}}+\frac{1}{\tau_l^2} A_le^{-\frac{t}{\tau_l}} \right).
\end{align}
The friction kernel $m$ in Eq.~\eqref{eq:Volterra},  is, as stated there, in general not given via a fluctuation function. As also noted above, for  systems with linear interactions, simplifications occur \cite{muller_properties_2020}, due to decoupling of averages and fluctuations. In this case, 
Eq.~\eqref{eq:FrictionKernelEq} is not only a mathematical, but also a physical inversion. Using   FDT for perturbations by a velocity \cite{hansen2013theory},  
\begin{align}
      \langle F(t) \rangle = - \beta \int_0^t dt' \langle F(t')F(0)\rangle_\eq v(t'). 
\end{align}
This can be used to identify the friction-kernel $m$ as 
\begin{align}
    m(t) &= \beta \gamma^{-1} \av{F(t) F(0)}_\eq - 2 \delta(t) \label{eq:mForceCorrelation}.
\end{align}
In contrast to Eq.~\eqref{eq:MobilityKernelEq}, the appearing force correlator is now measured with the probe particle {\it fixed} at a given position. This little detail changes the scenario from reciprocal to nonreciprocal, and the nonreciprocal timescales appear. Specifically, in the bath particle model, (see Appendix~\ref{chap:mIn2BP}),
\begin{align}
    m(t) = \frac{\kappa_s}{\gamma} \exp(-t/\tilde\tau_s) + \frac{\kappa_l}{\gamma} \exp(-t/\tilde\tau_l).\label{eq:ml}
\end{align}
The analysis of Eq.~\eqref{eq:Volterra} extends beyond the regime of linear interactions, where Eq.~\eqref{eq:ml} is not valid, but this discussion still provides useful intuition. 

With this in mind, we can now obtain insights via the Volterra relation between the two kernels,  Eq.~\eqref{eq:Volterra}.
With increasing concentration, the friction kernel at vanishing frequency, $s=0$,  viz.~$m(s=0)$ grows, which is a famous prediction of MCT. This can be directly seen from the following relation, which follows from Eq.~\eqref{eq:MSD_m}, 
\begin{align}
    D_\infty = \frac{D_0}{1+ \hat{m}(s=0)}.\label{eq:Dm}
\end{align}
Notably, in the bath particle model, as seen in Fig.~\ref{fig:TimeScales_RecoilAmplitudes}(a), the inverse long-time diffusion coefficient grows faster than the timescale   $\tilde\tau_l$. Indeed, it is also the amplitude of $m$ that grows with increasing density.

In contrast, Eq.~\eqref{eq:Volterra} shows that, if $m(s=0)$ grows, the mobility kernel $M(s=0)$ remains on order unity, hence does not grow with density. This is in agreement with the observation that the reciprocal timescales $\tau_{s,l}$ remain on the same order of magnitude.
This confirms the trends in Fig.~\ref{fig:TimeScales_RecoilAmplitudes}, and can also be seen in the expression for the reciprocal timescales $\tau_{s,l}$ (Eq.~\eqref{tauSLAnalytic}) which remains finite even in the limits $\tilde\tau_{s,l} \to \infty$.

Within MCT, increasing the concentration is modeled by an increase in the packing fraction. 
The "master recoil" in Fig.~\ref{fig:ShortScaling} then confirms the fast characteristics of the mobility kernel $M$; the formula is given in Eq.~\eqref{eq:recoil}. 
For comparison, we also show the corresponding friction-kernels $m(t)$ in the inset of Fig.~\ref{fig:ShortScaling}.
Clearly, $m(t)$ slows down with increasing packing fraction. 
As was already observed in Ref.~\onlinecite{Mandal2019} the kernel enters a regime of algebraic decay with a power of $-\frac{5}{2}$. For the lowest density shown this starts at about $t=\tau_B$ and thus after the corresponding duration of the recoil measurements.

\section{Conclusion}\label{chap:Conclusions}

Based on the Einstein relation generalized to the time-dependent linear response relation between the average position of a driven probe particle and its mean squared displacements in equilibrium, Eq.~\eqref{eq:LinearResponse}, we analyzed the response kernels of a viscoelastic system for a number of experimental densities. We investigated mean squared displacements (MSD) of colloidal probes in equilibrium, as well as recoils, when a driven probe reversed its motion and recovered parts of its displacement after the driving is turned off; the latter study was confined to the regime of small driving where linear response applies.

Theoretical analysis used a generic Langevin description where a small number of bath particles is coupled to the probe to model different relaxation channels, and the mode coupling theory (MCT) of structural dynamics in hard sphere solutions. Both descriptions showed that the slowing down  of the long time diffusion with increasing bath concentration is related to the friction kernel $m(t)$ that encodes the increasing viscoelastic memory in the surrounding bath (Eq.~\eqref{eq:Dm}). Its $s=0$ mode grows, as is often modelled by an ansatz following Maxwell $m(t)= f^{\rm MW}\,e^{-t/\tau^{\rm MW}}$, where, with Eq.~\eqref{eq:Dm}, the   long-time diffusion, $D_\infty^{-1}\sim f^{\rm MW} \tau^{\rm MW}$.  
Our theoretical models recover this interpretation of $D_\infty$, yet provide convincing evidence that a single-relaxation time ansatz following Maxwell is insufficient to understand the motion of a colloidal probe in a viscoelastic fluid. While this is visible in the MSD through the increasing window of sub-diffusive motion between short, $\lim_{t \to 0}\MSD (t) = 2 D_0 t$, and long time behaviors $\lim_{t \to \infty}\MSD (t) = 2 D_\infty t$, the mentioned timescales are however partly overshadowed by the long time diffusion process. This slow process is also difficult to measure in experiments, because of stability issues such as spatial drift or unsteady temperatures. This is in striking difference to recoils that are much simpler to measure.

The recoil curves clearly show distinct timescales, which are determined by a mobility kernel $M(t)$. $M(t)$ is  related to the friction kernel $m(t)$ (see Eq.~\eqref{eq:Volterra} in Laplace space) and is characterized by a different set of timescales, where one of them is far smaller than $\tau^{\rm MW}$. Recoil spectra measure the elastic response that, once its amplitude has developed in a viscoelastic fluid, exhibits a rather small temporal variation with concentration.

While, theoretically, all discussed kernels and observables are related, via Eqs.~\eqref{eq:RecoilMSDF} and \eqref{eq:Volterra}, this analysis of recoils and MSD in the same system revealed that measuring recoils provides several benefits: It allows access to the mobility kernel $M$ via Eq.~\eqref{eq:recoil}. This is easier than determining $M$ from  MSD data via Eq.~\eqref{eq:MMSD}, as the latter  requires accurate determination of long-time diffusion. The recoil also displays nontrivial timescales and relaxation channels, which in MSD can be hidden.

 Future work can address effects of nonlinearity, both in the driving speed, as well as in interactions.

\begin{acknowledgments}
This project was funded by the Deutsche Forschungsgemeinschaft (DFG), Grant No. SFB 1432 - Projects C05 and C06. F.G. acknowledges support by the Alexander von Humboldt foundation.
The authors declare no competing interest.
\end{acknowledgments}

\appendix 
\section{Experimental parameters}\label{chap:ExperimentalParameters}
In Table~\ref{tab:ExpParam} we list the trap stiffness $\kappa$ used for recoil experiments at different concentrations $c$. 
These values are obtained from fitting the equilibrium probability distribution of the trapped probe to a Boltzmann distribution $P(x) \propto e^{- U(x)/k_B T}$ with harmonic potential $U(x) = 1/2 \kappa x^2 $, see e.g.~Ref.~\onlinecite{Berner2018}.
\begin{table}[h]
\caption{Experimental trap stiffness $\kappa$ used for the recoil experiments.}
\label{tab:ExpParam}
\centering
\begin{ruledtabular}
\begin{tabular}{@{}cc@{}}
$c\,\SI{}{(\milli \Molar)}$           & $\kappa\,\SI{}{(\micro \N /\m)}$  \\ \hline
5 & $46 \pm 4$  \\
6 & $16 \pm 1$  \\
7 & $7.2 \pm 0.8$  \\
8 & -           \\
9 & -           \\
\end{tabular}
\end{ruledtabular}
\end{table}

\section{Model parameters}\label{chap:ModelParameters}

For the model parameters of the linear two-bath particle models fitting experimental MSD's and recoils see Table.~\ref{tab:MSD2BPFit}.
The model parameters for fitting hard sphere MCT to experimental data are listed in Table~\ref{tab:MSDMCTFit}.
\begin{table}[h]
\caption{Model parameters $\gamma$, $\gamma_s$, $\gamma_l$, $\kappa_s$ and $\kappa_l$ used for fitting a two-bath particle model to experimental MSD's and recoils for varying concentrations $c$.}
\label{tab:MSD2BPFit}
\begin{ruledtabular}
\begin{tabular}{@{}cccccc@{}}
$c$           & $\gamma$ & $\gamma_s$ & $\gamma_l$ & $\kappa_s$ & $\kappa_l$  \\ 
$\SI{}{(\milli \Molar)}$ & $\SI{}{(\micro \N \s /\m)}$ & $\SI{}{(\micro \N \s /\m)}$ & $\SI{}{(\micro \N \s /\m)}$ & $\SI{}{(\micro \N/\m)}$ & $\SI{}{(\micro \N /\m)}$ \\\hline
5 & 0.246                             & 0.272                               & 0.218                               & 0.37                             & 0.05                          \\
6 & 0.123 & 0.494 & 1.48 & 0.494 & 0.123   \\
7& 0.358 & 1.288 & 16.36 & 1.074 & 0.59                          \\
8 & 0.374 & 1.404 & 36 & 1.236 & 1.068                             \\
9 & 0.46 & 1.612 & 101 & 1.798 & 2.32\\

\end{tabular}
\end{ruledtabular}
\end{table}
\begin{table}[htb]
    \caption{Model parameters $\phi, \tau_B$ and $d^2$ for fitting hard sphere MCT to experimental MSD's and recoils for varying concentrations $c$. $D_0$ and $ D_\infty$ are the obtained short- and long-time diffusivities.}
    \label{tab:MSDMCTFit}
    \begin{ruledtabular}
    \begin{tabular}{@{}cccccc@{}}
        $c$                     & $\phi$    & $\tau_B$          &  $d^2$                    & $D_0$                             & $D_\infty$                        \\
          $\SI{}{(\milli\Molar)}$ & $\phi$    & $\SI{}{(\s)}$     &  $\SI{}{(\micro\m^2)}$    & $\SI{}{(10^{-2}\micro\m^2/s)}$    & $\SI{}{(10^{-4}\micro\m^2/s)}$    \\\hline
        5                       & 0.36      & 41.351            &  0.837                    & 2.024                             &  56.36                            \\
        6                       & 0.45      & 69.531            &  1.157                    & 1.664                             &  9.819                            \\
        7                       & 0.48      & 45.958            &  0.727                    & 1.581                             &  2.561                            \\
        8                       & 0.49      & 47.575            &  0.644                    & 1.353                             &  1.047                            \\
        9                       & 0.50      & 21.709            &  0.345                    & 1.589                             &  0.392                            \\ 
      
    \end{tabular}
    \end{ruledtabular}
\end{table}

\section{Recoil amplitudes within 2-BP model}\label{chap:RecoilAmplitudes2BP}

The recoil amplitudes in Eq.~\eqref{RecoilTheo} take the following form
\begin{align}
&\frac{A_{s}}{v} = \frac{\gamma_s\tilde\tau_s\left(1-e^{-\frac{t_\mathrm{sh}}{\tilde\tau_s}}\right)}{2 \left(\gamma+\gamma_s+\gamma_l\right)}\left[1+ \frac{\tilde\tau_l \left[\zeta_l (\tilde\tau_l-\tilde\tau_s)+\gamma_s+\gamma_l \right]}{(\gamma+\gamma_s+\gamma_l)(\tau_l-\tau_s)} \right] \label{eq:AslAnalytic} \notag\\
&+\frac{\gamma_l\tilde\tau_l\left(1-e^{-\frac{t_\mathrm{sh}}{\tilde\tau_l}}\right)}{2 \left(\gamma+\gamma_s+\gamma_l\right)}\left[1- \frac{\tilde\tau_s \left[\zeta_s (\tilde\tau_s-\tilde\tau_l)+\gamma_s+\gamma_l \right]}{(\gamma+\gamma_s+\gamma_l)(\tau_s-\tau_l)} \right],
\end{align}
$A_l$ follows from $A_s$ by changing indices $s\leftrightarrow l$ and  $A_\mathrm{tot}=A_s+A_l$~\cite{Ginot2022recoil}. 

\section{Alpha scaling of recoils}\label{chap:AlphaScaling}
 
\begin{figure}[h]
    \centering
    \includegraphics[]{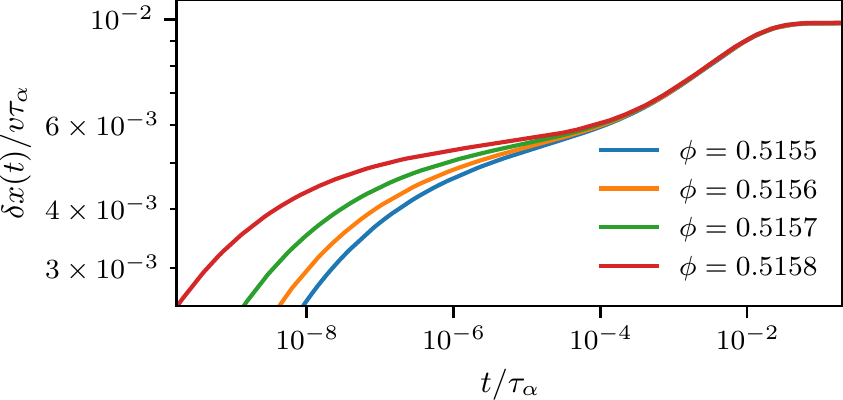}
    \caption{Recoils calculated from MCT MSD's via Eq.~\eqref{eq:RecoilMSDVNESS} and rescaled by the timescale $\tau_\alpha \propto D_\infty^{-1}$. Close to the glass transition ($\phi_c = 0.5159$), recoils approach a universal second relaxation step.}
    \label{fig:AlphaScaling}
\end{figure}
Figure \ref{fig:ShortScaling} 
shows that for increasing packing fraction the recoil curves exhibit a variation at long times. This is connected to the glass transition in the MCT hard sphere calculations. Calculating recoil curves in MCT for packing fractions very close to the glass transition we can observe that scaling both axes by the $\alpha$-time $ \tau_\alpha \propto D_\infty^{-1}$ leads to a universal description of a second relaxation step in the recoil. This can be seen in Fig.~\ref{fig:AlphaScaling}; the glass transition lies at $\phi_c=0.5159$ for the present numerics. The collapse on a master curve corresponds naturally to the typical $\alpha$-scaling relations considered for the hard sphere MSD before \cite{fuchs1998asymptotic}.

\section{Identification of $m(t)$ within the linear two-bath particle model}\label{chap:mIn2BP}

We can rewrite the set of overdamped Langevin equations, Eqs.~\eqref{eq:Langevin_tracer}-\eqref{eq:Langevin_noise}, describing the motion of the probe and the two bath particles in the linear two-bath particle model in terms of a generalized Langevin equation
\begin{align}
    \int_{-\infty}^t \mathrm{d}t' \, \Gamma(t-t') \dot{x}(t') = F_\mathrm{trap}(t) + \Tilde{\xi}(t)
    \label{eq:GLE}
\end{align}
with memory kernel
\begin{align}
    \Gamma(t) = 2 \gamma \delta(t) + \kappa_s e^{- \frac{\kappa_s}{\gamma_s}t} + \kappa_l e^{- \frac{\kappa_l}{\gamma_l}t}
    \label{eq:memory_kernel}
\end{align}
and noise
\begin{align}
    \Tilde{\xi}(t) = \xi(t) + \sum_{s,l} \frac{\kappa_n}{\gamma_n}\int_{-\infty}^t \mathrm{d}t' \, \xi_n(t') e^{-\frac{\kappa_n}{\gamma_n}(t-t')},
    \label{eq:General_Noise}
\end{align}
which fulfills FDT, $\langle \Tilde{\xi}(t) \Tilde{\xi}(t') \rangle = k_\mathrm{B}T \Gamma(|t-t'|)$.
This formulation allows us to find a relation between the velocity autocorrelation $C_{vv}(t) = \langle v(t) v(0)\rangle_\eq$ and the memory kernel $\Gamma(t)$, for which we take the bath to be prepared in equilibrium at $t=0$.
We start by multiplying Eq.~\eqref{eq:memory_kernel} in the absence of a trapping force, $F_\mathrm{trap}(t) = 0$ with $v(0)$ and performing an equilibrium average. This yields
\begin{align}
    \int_{0}^t dt' \Gamma(t-t') \langle v(t') v(0) \rangle_\eq = \langle \tilde\xi(t) v(0)\rangle_\eq.
\end{align}
Using $\langle \tilde\xi(t) v(0)\rangle_\eq = 2 k_B T$ we find in Laplace space
\begin{align}
    \hat{\Gamma}(s) \hat{C}_{vv}(s) = \frac{2 k_B T}{s}.
\end{align}
In order to identify the friction kernel $m(t)$ as defined for a microscopic system of spherical particles obeying Smoluchowski dynamics, we relate the velocity autocorrelation to the MSD (in Laplace space)
\begin{align}
    \frac{\hat{C}_{vv}}{s} = \frac{1}{2}s \hat{\MSD}(s),
\end{align}
which leads to
\begin{align}
    \hat{\MSD}(s) = \frac{2 k_B T}{s^2 \hat{\Gamma}(s)}.
    \label{eq:MSDGammaLaplace}
\end{align}
After transforming back to time space we obtain
\begin{align}
    \int_0^t dt' \Gamma(t-t') \MSD(t') = 2k_B T \cdot t.
    \label{eq:MSD_Gamma_EOM}
\end{align}
This expression can finally be compared to Eq.~\eqref{eq:MSD_m}, the equation of motion used in MCT.
We find $\Gamma(t) = \gamma m(t) + 2 \gamma \delta(t)$, as $D_0 = k_B T/\gamma$ in our model. Hence we can identify $m(t) = \frac{\kappa_s}{\gamma} e^{-\frac{\kappa_s}{\gamma_s}t}+ \frac{\kappa_l}{\gamma} e^{-\frac{\kappa_l}{\gamma_l}t}$.

\section*{Data availability}
The data that support the findings of this study are available
from the corresponding author upon reasonable request.

%


\section*{References}
\begin{thebibliography}{44}%
\makeatletter
\providecommand \@ifxundefined [1]{%
 \@ifx{#1\undefined}
}%
\providecommand \@ifnum [1]{%
 \ifnum #1\expandafter \@firstoftwo
 \else \expandafter \@secondoftwo
 \fi
}%
\providecommand \@ifx [1]{%
 \ifx #1\expandafter \@firstoftwo
 \else \expandafter \@secondoftwo
 \fi
}%
\providecommand \natexlab [1]{#1}%
\providecommand \enquote  [1]{``#1''}%
\providecommand \bibnamefont  [1]{#1}%
\providecommand \bibfnamefont [1]{#1}%
\providecommand \citenamefont [1]{#1}%
\providecommand \href@noop [0]{\@secondoftwo}%
\providecommand \href [0]{\begingroup \@sanitize@url \@href}%
\providecommand \@href[1]{\@@startlink{#1}\@@href}%
\providecommand \@@href[1]{\endgroup#1\@@endlink}%
\providecommand \@sanitize@url [0]{\catcode `\\12\catcode `\$12\catcode
  `\&12\catcode `\#12\catcode `\^12\catcode `\_12\catcode `\%12\relax}%
\providecommand \@@startlink[1]{}%
\providecommand \@@endlink[0]{}%
\providecommand \url  [0]{\begingroup\@sanitize@url \@url }%
\providecommand \@url [1]{\endgroup\@href {#1}{\urlprefix }}%
\providecommand \urlprefix  [0]{URL }%
\providecommand \Eprint [0]{\href }%
\providecommand \doibase [0]{http://dx.doi.org/}%
\providecommand \selectlanguage [0]{\@gobble}%
\providecommand \bibinfo  [0]{\@secondoftwo}%
\providecommand \bibfield  [0]{\@secondoftwo}%
\providecommand \translation [1]{[#1]}%
\providecommand \BibitemOpen [0]{}%
\providecommand \bibitemStop [0]{}%
\providecommand \bibitemNoStop [0]{.\EOS\space}%
\providecommand \EOS [0]{\spacefactor3000\relax}%
\providecommand \BibitemShut  [1]{\csname bibitem#1\endcsname}%
\let\auto@bib@innerbib\@empty
\bibitem [{\citenamefont {Larson}(1999)}]{larson_structure_1999}%
  \BibitemOpen
  \bibfield  {author} {\bibinfo {author} {\bibfnamefont {R.}~\bibnamefont
  {Larson}},\ }\href@noop {} {\emph {\bibinfo {title} {The {Structure} and
  {Rheology} of {Complex} {Fluids}}}}\ (\bibinfo  {publisher} {Oxford
  University Press},\ \bibinfo {year} {1999})\BibitemShut {NoStop}%
\bibitem [{\citenamefont {Squires}\ and\ \citenamefont
  {Mason}(2010)}]{squires2010fluid}%
  \BibitemOpen
  \bibfield  {author} {\bibinfo {author} {\bibfnamefont {T.~M.}\ \bibnamefont
  {Squires}}\ and\ \bibinfo {author} {\bibfnamefont {T.~G.}\ \bibnamefont
  {Mason}},\ }\href
  {https://www.annualreviews.org/doi/pdf/10.1146/annurev-fluid-121108-145608}
  {\bibfield  {journal} {\bibinfo  {journal} {Annu. Rev. Fluid Mech.}\ }\textbf
  {\bibinfo {volume} {42}},\ \bibinfo {pages} {413} (\bibinfo {year}
  {2010})}\BibitemShut {NoStop}%
\bibitem [{\citenamefont {Puertas}\ and\ \citenamefont
  {Voigtmann}(2014)}]{puertas_microrheology_2014}%
  \BibitemOpen
  \bibfield  {author} {\bibinfo {author} {\bibfnamefont {A.~M.}\ \bibnamefont
  {Puertas}}\ and\ \bibinfo {author} {\bibfnamefont {T.}~\bibnamefont
  {Voigtmann}},\ }\href {\doibase 10.1088/0953-8984/26/24/243101} {\bibfield
  {journal} {\bibinfo  {journal} {J. Phys.: Condens. Matter}\ }\textbf
  {\bibinfo {volume} {26}},\ \bibinfo {pages} {243101} (\bibinfo {year}
  {2014})}\BibitemShut {NoStop}%
\bibitem [{\citenamefont {Furst}\ and\ \citenamefont
  {Squires}(2017)}]{furst_microrheology_2017}%
  \BibitemOpen
  \bibfield  {author} {\bibinfo {author} {\bibfnamefont {E.}~\bibnamefont
  {Furst}}\ and\ \bibinfo {author} {\bibfnamefont {T.}~\bibnamefont
  {Squires}},\ }\href@noop {} {\emph {\bibinfo {title} {Microrheology}}}\
  (\bibinfo  {publisher} {Oxford University Press},\ \bibinfo {year}
  {2017})\BibitemShut {NoStop}%
\bibitem [{\citenamefont {Gomez-Solano}\ and\ \citenamefont
  {Bechinger}(2014)}]{gomez-solano_probing_2014}%
  \BibitemOpen
  \bibfield  {author} {\bibinfo {author} {\bibfnamefont {J.~R.}\ \bibnamefont
  {Gomez-Solano}}\ and\ \bibinfo {author} {\bibfnamefont {C.}~\bibnamefont
  {Bechinger}},\ }\href {\doibase 10.1209/0295-5075/108/54008} {\bibfield
  {journal} {\bibinfo  {journal} {Europhys. Lett.}\ }\textbf {\bibinfo {volume}
  {108}},\ \bibinfo {pages} {54008} (\bibinfo {year} {2014})}\BibitemShut
  {NoStop}%
\bibitem [{\citenamefont {Jain}\ \emph {et~al.}(2021)\citenamefont {Jain},
  \citenamefont {Ginot}, \citenamefont {Berner}, \citenamefont {Bechinger},\
  and\ \citenamefont {Kr{\"u}ger}}]{jain2021two}%
  \BibitemOpen
  \bibfield  {author} {\bibinfo {author} {\bibfnamefont {R.}~\bibnamefont
  {Jain}}, \bibinfo {author} {\bibfnamefont {F.}~\bibnamefont {Ginot}},
  \bibinfo {author} {\bibfnamefont {J.}~\bibnamefont {Berner}}, \bibinfo
  {author} {\bibfnamefont {C.}~\bibnamefont {Bechinger}}, \ and\ \bibinfo
  {author} {\bibfnamefont {M.}~\bibnamefont {Kr{\"u}ger}},\ }\href
  {https://aip.scitation.org/doi/pdf/10.1063/5.0048320} {\bibfield  {journal}
  {\bibinfo  {journal} {J. Chem. Phys.}\ }\textbf {\bibinfo {volume} {154}},\
  \bibinfo {pages} {184904} (\bibinfo {year} {2021})}\BibitemShut {NoStop}%
\bibitem [{\citenamefont {Berner}\ \emph {et~al.}(2018)\citenamefont {Berner},
  \citenamefont {M{\"u}ller}, \citenamefont {Gomez-Solano}, \citenamefont
  {Kr{\"u}ger},\ and\ \citenamefont {Bechinger}}]{Berner2018}%
  \BibitemOpen
  \bibfield  {author} {\bibinfo {author} {\bibfnamefont {J.}~\bibnamefont
  {Berner}}, \bibinfo {author} {\bibfnamefont {B.}~\bibnamefont {M{\"u}ller}},
  \bibinfo {author} {\bibfnamefont {J.~R.}\ \bibnamefont {Gomez-Solano}},
  \bibinfo {author} {\bibfnamefont {M.}~\bibnamefont {Kr{\"u}ger}}, \ and\
  \bibinfo {author} {\bibfnamefont {C.}~\bibnamefont {Bechinger}},\ }\href
  {https://www.nature.com/articles/s41467-018-03345-2} {\bibfield  {journal}
  {\bibinfo  {journal} {Nature commun.}\ }\textbf {\bibinfo {volume} {9}},\
  \bibinfo {pages} {1} (\bibinfo {year} {2018})}\BibitemShut {NoStop}%
\bibitem [{\citenamefont {Jayaraman}\ and\ \citenamefont
  {Belmonte}(2003)}]{jayaraman_oscillations_2003}%
  \BibitemOpen
  \bibfield  {author} {\bibinfo {author} {\bibfnamefont {A.}~\bibnamefont
  {Jayaraman}}\ and\ \bibinfo {author} {\bibfnamefont {A.}~\bibnamefont
  {Belmonte}},\ }\href {\doibase 10.1103/PhysRevE.67.065301} {\bibfield
  {journal} {\bibinfo  {journal} {Phys. Rev. E}\ }\textbf {\bibinfo {volume}
  {67}},\ \bibinfo {pages} {065301} (\bibinfo {year} {2003})}\BibitemShut
  {NoStop}%
\bibitem [{\citenamefont {Handzy}\ and\ \citenamefont
  {Belmonte}(2004)}]{handzy_oscillatory_2004}%
  \BibitemOpen
  \bibfield  {author} {\bibinfo {author} {\bibfnamefont {N.~Z.}\ \bibnamefont
  {Handzy}}\ and\ \bibinfo {author} {\bibfnamefont {A.}~\bibnamefont
  {Belmonte}},\ }\href {\doibase 10.1103/PhysRevLett.92.124501} {\bibfield
  {journal} {\bibinfo  {journal} {Phys. Rev. Lett.}\ }\textbf {\bibinfo
  {volume} {92}},\ \bibinfo {pages} {124501} (\bibinfo {year}
  {2004})}\BibitemShut {NoStop}%
\bibitem [{\citenamefont {Gomez-Solano}\ and\ \citenamefont
  {Bechinger}(2015)}]{Gomez-Solano2015-qu}%
  \BibitemOpen
  \bibfield  {author} {\bibinfo {author} {\bibfnamefont {J.~R.}\ \bibnamefont
  {Gomez-Solano}}\ and\ \bibinfo {author} {\bibfnamefont {C.}~\bibnamefont
  {Bechinger}},\ }\href
  {https://iopscience.iop.org/article/10.1088/1367-2630/17/10/103032/pdf}
  {\bibfield  {journal} {\bibinfo  {journal} {New J. Phys.}\ }\textbf {\bibinfo
  {volume} {17}},\ \bibinfo {pages} {103032} (\bibinfo {year}
  {2015})}\BibitemShut {NoStop}%
\bibitem [{\citenamefont {Ginot}\ \emph
  {et~al.}(2022{\natexlab{a}})\citenamefont {Ginot}, \citenamefont {Caspers},
  \citenamefont {Reinalter}, \citenamefont {Kumar}, \citenamefont {Krüger},\
  and\ \citenamefont {Bechinger}}]{Ginot2022recoil}%
  \BibitemOpen
  \bibfield  {author} {\bibinfo {author} {\bibfnamefont {F.}~\bibnamefont
  {Ginot}}, \bibinfo {author} {\bibfnamefont {J.}~\bibnamefont {Caspers}},
  \bibinfo {author} {\bibfnamefont {L.~F.}\ \bibnamefont {Reinalter}}, \bibinfo
  {author} {\bibfnamefont {K.~K.}\ \bibnamefont {Kumar}}, \bibinfo {author}
  {\bibfnamefont {M.}~\bibnamefont {Krüger}}, \ and\ \bibinfo {author}
  {\bibfnamefont {C.}~\bibnamefont {Bechinger}},\ }\href
  {https://arxiv.org/abs/2204.02369} {\bibfield  {journal} {\bibinfo  {journal}
  {arXiv:2204.02369}\ } (\bibinfo {year} {2022}{\natexlab{a}})}\BibitemShut
  {NoStop}%
\bibitem [{\citenamefont {Wilson}\ \emph {et~al.}(2011)\citenamefont {Wilson},
  \citenamefont {Harrison}, \citenamefont {Poon},\ and\ \citenamefont
  {Puertas}}]{wilson_microrheology_2011}%
  \BibitemOpen
  \bibfield  {author} {\bibinfo {author} {\bibfnamefont {L.~G.}\ \bibnamefont
  {Wilson}}, \bibinfo {author} {\bibfnamefont {A.~W.}\ \bibnamefont
  {Harrison}}, \bibinfo {author} {\bibfnamefont {W.~C.~K.}\ \bibnamefont
  {Poon}}, \ and\ \bibinfo {author} {\bibfnamefont {A.~M.}\ \bibnamefont
  {Puertas}},\ }\href {\doibase 10.1209/0295-5075/93/58007} {\bibfield
  {journal} {\bibinfo  {journal} {Europhys. Lett.}\ }\textbf {\bibinfo {volume}
  {93}},\ \bibinfo {pages} {58007} (\bibinfo {year} {2011})}\BibitemShut
  {NoStop}%
\bibitem [{\citenamefont {Harrer}\ \emph {et~al.}(2012)\citenamefont {Harrer},
  \citenamefont {Winter}, \citenamefont {Horbach}, \citenamefont {Fuchs},\ and\
  \citenamefont {Voigtmann}}]{harrer_force-induced_2012}%
  \BibitemOpen
  \bibfield  {author} {\bibinfo {author} {\bibfnamefont {C.~J.}\ \bibnamefont
  {Harrer}}, \bibinfo {author} {\bibfnamefont {D.}~\bibnamefont {Winter}},
  \bibinfo {author} {\bibfnamefont {J.}~\bibnamefont {Horbach}}, \bibinfo
  {author} {\bibfnamefont {M.}~\bibnamefont {Fuchs}}, \ and\ \bibinfo {author}
  {\bibfnamefont {T.}~\bibnamefont {Voigtmann}},\ }\href {\doibase
  10.1088/0953-8984/24/46/464105} {\bibfield  {journal} {\bibinfo  {journal}
  {J. Phys.: Condens. Matter}\ }\textbf {\bibinfo {volume} {24}},\ \bibinfo
  {pages} {464105} (\bibinfo {year} {2012})}\BibitemShut {NoStop}%
\bibitem [{\citenamefont {Gazuz}\ \emph {et~al.}(2009)\citenamefont {Gazuz},
  \citenamefont {Puertas}, \citenamefont {Voigtmann},\ and\ \citenamefont
  {Fuchs}}]{gazuz_active_2009}%
  \BibitemOpen
  \bibfield  {author} {\bibinfo {author} {\bibfnamefont {I.}~\bibnamefont
  {Gazuz}}, \bibinfo {author} {\bibfnamefont {A.~M.}\ \bibnamefont {Puertas}},
  \bibinfo {author} {\bibfnamefont {T.}~\bibnamefont {Voigtmann}}, \ and\
  \bibinfo {author} {\bibfnamefont {M.}~\bibnamefont {Fuchs}},\ }\href
  {\doibase 10.1103/PhysRevLett.102.248302} {\bibfield  {journal} {\bibinfo
  {journal} {Phys. Rev. Lett.}\ }\textbf {\bibinfo {volume} {102}},\ \bibinfo
  {pages} {248302} (\bibinfo {year} {2009})}\BibitemShut {NoStop}%
\bibitem [{\citenamefont {Squires}\ and\ \citenamefont
  {Brady}(2005)}]{Squires2005}%
  \BibitemOpen
  \bibfield  {author} {\bibinfo {author} {\bibfnamefont {T.~M.}\ \bibnamefont
  {Squires}}\ and\ \bibinfo {author} {\bibfnamefont {J.~F.}\ \bibnamefont
  {Brady}},\ }\href {\doibase 10.1063/1.1960607} {\bibfield  {journal}
  {\bibinfo  {journal} {Phys. Fluids}\ }\textbf {\bibinfo {volume} {17}},\
  \bibinfo {pages} {073101} (\bibinfo {year} {2005})}\BibitemShut {NoStop}%
\bibitem [{\citenamefont {Şenbil}\ \emph {et~al.}(2019)\citenamefont
  {Şenbil}, \citenamefont {Gruber}, \citenamefont {Zhang}, \citenamefont
  {Fuchs},\ and\ \citenamefont {Scheffold}}]{senbil_observation_2019}%
  \BibitemOpen
  \bibfield  {author} {\bibinfo {author} {\bibfnamefont {N.}~\bibnamefont
  {Şenbil}}, \bibinfo {author} {\bibfnamefont {M.}~\bibnamefont {Gruber}},
  \bibinfo {author} {\bibfnamefont {C.}~\bibnamefont {Zhang}}, \bibinfo
  {author} {\bibfnamefont {M.}~\bibnamefont {Fuchs}}, \ and\ \bibinfo {author}
  {\bibfnamefont {F.}~\bibnamefont {Scheffold}},\ }\href {\doibase
  10.1103/PhysRevLett.122.108002} {\bibfield  {journal} {\bibinfo  {journal}
  {Phys. Rev. Lett.}\ }\textbf {\bibinfo {volume} {122}},\ \bibinfo {pages}
  {108002} (\bibinfo {year} {2019})}\BibitemShut {NoStop}%
\bibitem [{\citenamefont {Dhont}(1996)}]{Dhont}%
  \BibitemOpen
  \bibfield  {author} {\bibinfo {author} {\bibfnamefont {J.~K.~G.}\
  \bibnamefont {Dhont}},\ }\href
  {https://www.elsevier.com/books/an-introduction-to-dynamics-of-colloids/dhont/978-0-444-82009-9}
  {\emph {\bibinfo {title} {An Introduction to Dynamics of Colloids}}}\
  (\bibinfo  {publisher} {Elsevier},\ \bibinfo {address} {Amsterdam},\ \bibinfo
  {year} {1996})\BibitemShut {NoStop}%
\bibitem [{\citenamefont {Hansen}\ and\ \citenamefont
  {McDonald}(2013)}]{hansen2013theory}%
  \BibitemOpen
  \bibfield  {author} {\bibinfo {author} {\bibfnamefont {J.~P.}\ \bibnamefont
  {Hansen}}\ and\ \bibinfo {author} {\bibfnamefont {I.~R.}\ \bibnamefont
  {McDonald}},\ }\href@noop {} {\emph {\bibinfo {title} {Theory of Simple
  Liquids: with Applications to Soft Matter}}}\ (\bibinfo  {publisher}
  {Elsevier Science},\ \bibinfo {year} {2013})\BibitemShut {NoStop}%
\bibitem [{\citenamefont {G{\"o}tze}(2009)}]{gotze2009complex}%
  \BibitemOpen
  \bibfield  {author} {\bibinfo {author} {\bibfnamefont {W.}~\bibnamefont
  {G{\"o}tze}},\ }\href@noop {} {\emph {\bibinfo {title} {Complex dynamics of
  glass-forming liquids: A mode-coupling theory}}},\ Vol.\ \bibinfo {volume}
  {143}\ (\bibinfo  {publisher} {Oxford University Press},\ \bibinfo {year}
  {2009})\BibitemShut {NoStop}%
\bibitem [{\citenamefont {Saito}\ and\ \citenamefont
  {Sakaue}(2017)}]{saito_complementary_2017}%
  \BibitemOpen
  \bibfield  {author} {\bibinfo {author} {\bibfnamefont {T.}~\bibnamefont
  {Saito}}\ and\ \bibinfo {author} {\bibfnamefont {T.}~\bibnamefont {Sakaue}},\
  }\href {\doibase 10.1103/PhysRevE.95.042143} {\bibfield  {journal} {\bibinfo
  {journal} {Phys. Rev. E}\ }\textbf {\bibinfo {volume} {95}},\ \bibinfo
  {pages} {042143} (\bibinfo {year} {2017})}\BibitemShut {NoStop}%
\bibitem [{\citenamefont {Daldrop}\ \emph {et~al.}(2017)\citenamefont
  {Daldrop}, \citenamefont {Kowalik},\ and\ \citenamefont
  {Netz}}]{daldrop_external_2017}%
  \BibitemOpen
  \bibfield  {author} {\bibinfo {author} {\bibfnamefont {J.~O.}\ \bibnamefont
  {Daldrop}}, \bibinfo {author} {\bibfnamefont {B.~G.}\ \bibnamefont
  {Kowalik}}, \ and\ \bibinfo {author} {\bibfnamefont {R.~R.}\ \bibnamefont
  {Netz}},\ }\href {\doibase 10.1103/PhysRevX.7.041065} {\bibfield  {journal}
  {\bibinfo  {journal} {Phys. Rev. X}\ }\textbf {\bibinfo {volume} {7}},\
  \bibinfo {pages} {041065} (\bibinfo {year} {2017})}\BibitemShut {NoStop}%
\bibitem [{\citenamefont {Kowalik}\ \emph {et~al.}(2019)\citenamefont
  {Kowalik}, \citenamefont {Daldrop}, \citenamefont {Kappler}, \citenamefont
  {Schulz}, \citenamefont {Schlaich},\ and\ \citenamefont
  {Netz}}]{Kowalik2019}%
  \BibitemOpen
  \bibfield  {author} {\bibinfo {author} {\bibfnamefont {B.}~\bibnamefont
  {Kowalik}}, \bibinfo {author} {\bibfnamefont {J.~O.}\ \bibnamefont
  {Daldrop}}, \bibinfo {author} {\bibfnamefont {J.}~\bibnamefont {Kappler}},
  \bibinfo {author} {\bibfnamefont {J.~C.~F.}\ \bibnamefont {Schulz}}, \bibinfo
  {author} {\bibfnamefont {A.}~\bibnamefont {Schlaich}}, \ and\ \bibinfo
  {author} {\bibfnamefont {R.~R.}\ \bibnamefont {Netz}},\ }\href {\doibase
  10.1103/PhysRevE.100.012126} {\bibfield  {journal} {\bibinfo  {journal}
  {Phys. Rev. E}\ }\textbf {\bibinfo {volume} {100}},\ \bibinfo {pages}
  {012126} (\bibinfo {year} {2019})}\BibitemShut {NoStop}%
\bibitem [{\citenamefont {Müller}\ \emph {et~al.}(2020)\citenamefont
  {Müller}, \citenamefont {Berner}, \citenamefont {Bechinger},\ and\
  \citenamefont {Krüger}}]{muller_properties_2020}%
  \BibitemOpen
  \bibfield  {author} {\bibinfo {author} {\bibfnamefont {B.}~\bibnamefont
  {Müller}}, \bibinfo {author} {\bibfnamefont {J.}~\bibnamefont {Berner}},
  \bibinfo {author} {\bibfnamefont {C.}~\bibnamefont {Bechinger}}, \ and\
  \bibinfo {author} {\bibfnamefont {M.}~\bibnamefont {Krüger}},\ }\href
  {\doibase 10.1088/1367-2630/ab6a39} {\bibfield  {journal} {\bibinfo
  {journal} {New J. Phys.}\ }\textbf {\bibinfo {volume} {22}},\ \bibinfo
  {pages} {023014} (\bibinfo {year} {2020})}\BibitemShut {NoStop}%
\bibitem [{\citenamefont {Basu}\ \emph {et~al.}(2022)\citenamefont {Basu},
  \citenamefont {Démery},\ and\ \citenamefont
  {Gambassi}}]{basu_dynamics_2022}%
  \BibitemOpen
  \bibfield  {author} {\bibinfo {author} {\bibfnamefont {U.}~\bibnamefont
  {Basu}}, \bibinfo {author} {\bibfnamefont {V.}~\bibnamefont {Démery}}, \
  and\ \bibinfo {author} {\bibfnamefont {A.}~\bibnamefont {Gambassi}},\ }\href
  {https://arxiv.org/abs/2203.13702} {\bibfield  {journal} {\bibinfo  {journal}
  {arXiv:2203.13702}\ } (\bibinfo {year} {2022})}\BibitemShut {NoStop}%
\bibitem [{\citenamefont {Rehage}\ and\ \citenamefont
  {Hoffmann}(1988)}]{Rehage1988}%
  \BibitemOpen
  \bibfield  {author} {\bibinfo {author} {\bibfnamefont {H.}~\bibnamefont
  {Rehage}}\ and\ \bibinfo {author} {\bibfnamefont {H.}~\bibnamefont
  {Hoffmann}},\ }\href {\doibase 10.1021/j100327a031} {\bibfield  {journal}
  {\bibinfo  {journal} {J. Phys. Chem.}\ }\textbf {\bibinfo {volume} {92}},\
  \bibinfo {pages} {4712} (\bibinfo {year} {1988})}\BibitemShut {NoStop}%
\bibitem [{\citenamefont {Spenley}\ \emph {et~al.}(1993)\citenamefont
  {Spenley}, \citenamefont {Cates},\ and\ \citenamefont
  {McLeish}}]{Spenley1993}%
  \BibitemOpen
  \bibfield  {author} {\bibinfo {author} {\bibfnamefont {N.~A.}\ \bibnamefont
  {Spenley}}, \bibinfo {author} {\bibfnamefont {M.~E.}\ \bibnamefont {Cates}},
  \ and\ \bibinfo {author} {\bibfnamefont {T.~C.~B.}\ \bibnamefont {McLeish}},\
  }\href {\doibase 10.1103/PhysRevLett.71.939} {\bibfield  {journal} {\bibinfo
  {journal} {Phys. Rev. Lett.}\ }\textbf {\bibinfo {volume} {71}},\ \bibinfo
  {pages} {939} (\bibinfo {year} {1993})}\BibitemShut {NoStop}%
\bibitem [{\citenamefont {Berret}\ \emph {et~al.}(1997)\citenamefont {Berret},
  \citenamefont {Porte},\ and\ \citenamefont {Decruppe}}]{Berret1997}%
  \BibitemOpen
  \bibfield  {author} {\bibinfo {author} {\bibfnamefont {J.-F.}\ \bibnamefont
  {Berret}}, \bibinfo {author} {\bibfnamefont {G.}~\bibnamefont {Porte}}, \
  and\ \bibinfo {author} {\bibfnamefont {J.-P.}\ \bibnamefont {Decruppe}},\
  }\href {\doibase 10.1103/PhysRevE.55.1668} {\bibfield  {journal} {\bibinfo
  {journal} {Phys. Rev. E}\ }\textbf {\bibinfo {volume} {55}},\ \bibinfo
  {pages} {1668} (\bibinfo {year} {1997})}\BibitemShut {NoStop}%
\bibitem [{\citenamefont {Cates}\ and\ \citenamefont
  {Candau}(1990)}]{cates1990statics}%
  \BibitemOpen
  \bibfield  {author} {\bibinfo {author} {\bibfnamefont {M.~E.}\ \bibnamefont
  {Cates}}\ and\ \bibinfo {author} {\bibfnamefont {S.~J.}\ \bibnamefont
  {Candau}},\ }\href
  {https://iopscience.iop.org/article/10.1088/0953-8984/2/33/001/pdf}
  {\bibfield  {journal} {\bibinfo  {journal} {J. Phys. Condens. Matter}\
  }\textbf {\bibinfo {volume} {2}},\ \bibinfo {pages} {6869} (\bibinfo {year}
  {1990})}\BibitemShut {NoStop}%
\bibitem [{\citenamefont {Jeon}\ \emph {et~al.}(2013)\citenamefont {Jeon},
  \citenamefont {Leijnse}, \citenamefont {Oddershede},\ and\ \citenamefont
  {Metzler}}]{Jeon_2013}%
  \BibitemOpen
  \bibfield  {author} {\bibinfo {author} {\bibfnamefont {J.-H.}\ \bibnamefont
  {Jeon}}, \bibinfo {author} {\bibfnamefont {N.}~\bibnamefont {Leijnse}},
  \bibinfo {author} {\bibfnamefont {L.~B.}\ \bibnamefont {Oddershede}}, \ and\
  \bibinfo {author} {\bibfnamefont {R.}~\bibnamefont {Metzler}},\ }\href
  {\doibase 10.1088/1367-2630/15/4/045011} {\bibfield  {journal} {\bibinfo
  {journal} {New J. Phys.}\ }\textbf {\bibinfo {volume} {15}},\ \bibinfo
  {pages} {045011} (\bibinfo {year} {2013})}\BibitemShut {NoStop}%
\bibitem [{\citenamefont {Baiesi}\ \emph {et~al.}(2021)\citenamefont {Baiesi},
  \citenamefont {Iubini},\ and\ \citenamefont {Orlandini}}]{baiesi2021}%
  \BibitemOpen
  \bibfield  {author} {\bibinfo {author} {\bibfnamefont {M.}~\bibnamefont
  {Baiesi}}, \bibinfo {author} {\bibfnamefont {S.}~\bibnamefont {Iubini}}, \
  and\ \bibinfo {author} {\bibfnamefont {E.}~\bibnamefont {Orlandini}},\ }\href
  {\doibase 10.1063/5.0072374} {\bibfield  {journal} {\bibinfo  {journal} {J.
  Chem. Phys.}\ }\textbf {\bibinfo {volume} {155}},\ \bibinfo {pages} {214905}
  (\bibinfo {year} {2021})}\BibitemShut {NoStop}%
\bibitem [{\citenamefont {Ginot}\ \emph
  {et~al.}(2022{\natexlab{b}})\citenamefont {Ginot}, \citenamefont {Caspers},
  \citenamefont {Kr\"uger},\ and\ \citenamefont {Bechinger}}]{Ginot2022}%
  \BibitemOpen
  \bibfield  {author} {\bibinfo {author} {\bibfnamefont {F.}~\bibnamefont
  {Ginot}}, \bibinfo {author} {\bibfnamefont {J.}~\bibnamefont {Caspers}},
  \bibinfo {author} {\bibfnamefont {M.}~\bibnamefont {Kr\"uger}}, \ and\
  \bibinfo {author} {\bibfnamefont {C.}~\bibnamefont {Bechinger}},\ }\href
  {\doibase 10.1103/PhysRevLett.128.028001} {\bibfield  {journal} {\bibinfo
  {journal} {Phys. Rev. Lett.}\ }\textbf {\bibinfo {volume} {128}},\ \bibinfo
  {pages} {028001} (\bibinfo {year} {2022}{\natexlab{b}})}\BibitemShut
  {NoStop}%
\bibitem [{\citenamefont {van Zanten}\ and\ \citenamefont
  {Rufener}(2000)}]{van_zanten_brownian_2000}%
  \BibitemOpen
  \bibfield  {author} {\bibinfo {author} {\bibfnamefont {J.~H.}\ \bibnamefont
  {van Zanten}}\ and\ \bibinfo {author} {\bibfnamefont {K.~P.}\ \bibnamefont
  {Rufener}},\ }\href {\doibase 10.1103/PhysRevE.62.5389} {\bibfield  {journal}
  {\bibinfo  {journal} {Phys. Rev. E}\ }\textbf {\bibinfo {volume} {62}},\
  \bibinfo {pages} {5389} (\bibinfo {year} {2000})}\BibitemShut {NoStop}%
\bibitem [{\citenamefont {Lu}\ and\ \citenamefont
  {Solomon}(2002)}]{lu_probe_2002}%
  \BibitemOpen
  \bibfield  {author} {\bibinfo {author} {\bibfnamefont {Q.}~\bibnamefont
  {Lu}}\ and\ \bibinfo {author} {\bibfnamefont {M.~J.}\ \bibnamefont
  {Solomon}},\ }\href {\doibase 10.1103/PhysRevE.66.061504} {\bibfield
  {journal} {\bibinfo  {journal} {Phys. Rev. E}\ }\textbf {\bibinfo {volume}
  {66}},\ \bibinfo {pages} {061504} (\bibinfo {year} {2002})}\BibitemShut
  {NoStop}%
\bibitem [{\citenamefont {van~der Gucht}\ \emph {et~al.}(2003)\citenamefont
  {van~der Gucht}, \citenamefont {Besseling}, \citenamefont {Knoben},
  \citenamefont {Bouteiller},\ and\ \citenamefont
  {Cohen~Stuart}}]{vandergucht2003}%
  \BibitemOpen
  \bibfield  {author} {\bibinfo {author} {\bibfnamefont {J.}~\bibnamefont
  {van~der Gucht}}, \bibinfo {author} {\bibfnamefont {N.~A.~M.}\ \bibnamefont
  {Besseling}}, \bibinfo {author} {\bibfnamefont {W.}~\bibnamefont {Knoben}},
  \bibinfo {author} {\bibfnamefont {L.}~\bibnamefont {Bouteiller}}, \ and\
  \bibinfo {author} {\bibfnamefont {M.~A.}\ \bibnamefont {Cohen~Stuart}},\
  }\href {\doibase 10.1103/PhysRevE.67.051106} {\bibfield  {journal} {\bibinfo
  {journal} {Phys. Rev. E}\ }\textbf {\bibinfo {volume} {67}},\ \bibinfo
  {pages} {051106} (\bibinfo {year} {2003})}\BibitemShut {NoStop}%
\bibitem [{\citenamefont {Nägele}(1996)}]{NAGELE1996215}%
  \BibitemOpen
  \bibfield  {author} {\bibinfo {author} {\bibfnamefont {G.}~\bibnamefont
  {Nägele}},\ }\href {\doibase https://doi.org/10.1016/0370-1573(95)00078-X}
  {\bibfield  {journal} {\bibinfo  {journal} {Phys. Rep.}\ }\textbf {\bibinfo
  {volume} {272}},\ \bibinfo {pages} {215} (\bibinfo {year}
  {1996})}\BibitemShut {NoStop}%
\bibitem [{\citenamefont {Crocker}\ and\ \citenamefont
  {Grier}(1996)}]{crocker1996methods}%
  \BibitemOpen
  \bibfield  {author} {\bibinfo {author} {\bibfnamefont {J.~C.}\ \bibnamefont
  {Crocker}}\ and\ \bibinfo {author} {\bibfnamefont {D.~G.}\ \bibnamefont
  {Grier}},\ }\href
  {https://d1wqtxts1xzle7.cloudfront.net/39887646/CrockerGrier1996b-with-cover-page-v2.pdf?Expires=1648812687&Signature=Ts1D~KWO1v4z9iMVwTDVa36bYlkm9Htafa3TEiI-C1jJznT7jq2oiLM8OZPxQhKgKWGn4xzH4pgAHnwuIG7PMd3fHer2WmWWLWo2tqQZ4yQ7CDTq78p14N0QyO3gKiYLjooRB9z94B5l5IBijjEUaG~5QZb3Eu3kETd~KtNhCCbO2owwt70k696bvJs8VnT8NECgFRx5f~Yb0UrNtdTaH1HKiOVDGU22Yi88tDIBsCGrdUPYa630RKeb0B4bj2y7MReHP6YlXVdAmaDYI2ZSq8BONCQ4Q3RfztXZtQOdq9q~VNchDA3ElDzhLQsIb61pDrA-VUBB-wNJqNDT78X60Q__&Key-Pair-Id=APKAJLOHF5GGSLRBV4ZA}
  {\bibfield  {journal} {\bibinfo  {journal} {J. Colloid Interface Sci.}\
  }\textbf {\bibinfo {volume} {179}},\ \bibinfo {pages} {298} (\bibinfo {year}
  {1996})}\BibitemShut {NoStop}%
\bibitem [{\citenamefont {Doerries}\ \emph {et~al.}(2021)\citenamefont
  {Doerries}, \citenamefont {Loos},\ and\ \citenamefont
  {Klapp}}]{doerries2021correlation}%
  \BibitemOpen
  \bibfield  {author} {\bibinfo {author} {\bibfnamefont {T.~J.}\ \bibnamefont
  {Doerries}}, \bibinfo {author} {\bibfnamefont {S.~A.~M.}\ \bibnamefont
  {Loos}}, \ and\ \bibinfo {author} {\bibfnamefont {S.~H.~L.}\ \bibnamefont
  {Klapp}},\ }\href {https://doi.org/10.1088/1742-5468/abdead} {\bibfield
  {journal} {\bibinfo  {journal} {J. Stat. Mech. Theor. Exp.}\ }\textbf
  {\bibinfo {volume} {2021}},\ \bibinfo {pages} {033202} (\bibinfo {year}
  {2021})}\BibitemShut {NoStop}%
\bibitem [{\citenamefont {Cichocki}\ and\ \citenamefont
  {Hess}(1987)}]{CichockiHess}%
  \BibitemOpen
  \bibfield  {author} {\bibinfo {author} {\bibfnamefont {B.}~\bibnamefont
  {Cichocki}}\ and\ \bibinfo {author} {\bibfnamefont {W.}~\bibnamefont
  {Hess}},\ }\href {https://aip.scitation.org/doi/10.1063/1.464523} {\bibfield
  {journal} {\bibinfo  {journal} {Physica A}\ }\textbf {\bibinfo {volume}
  {141}},\ \bibinfo {pages} {475} (\bibinfo {year} {1987})}\BibitemShut
  {NoStop}%
\bibitem [{\citenamefont {Gazuz}\ and\ \citenamefont
  {Fuchs}(2013)}]{Gazuz2013}%
  \BibitemOpen
  \bibfield  {author} {\bibinfo {author} {\bibfnamefont {I.}~\bibnamefont
  {Gazuz}}\ and\ \bibinfo {author} {\bibfnamefont {M.}~\bibnamefont {Fuchs}},\
  }\href {\doibase 10.1103/PhysRevE.87.032304} {\bibfield  {journal} {\bibinfo
  {journal} {Phys. Rev. E}\ }\textbf {\bibinfo {volume} {87}},\ \bibinfo
  {pages} {032304} (\bibinfo {year} {2013})}\BibitemShut {NoStop}%
\bibitem [{\citenamefont {Tricomi}(1985)}]{tricomi_integral_1985}%
  \BibitemOpen
  \bibfield  {author} {\bibinfo {author} {\bibfnamefont {F.}~\bibnamefont
  {Tricomi}},\ }\href@noop {} {\emph {\bibinfo {title} {Integral
  {Equations}}}}\ (\bibinfo  {publisher} {Dover Publications},\ \bibinfo {year}
  {1985})\BibitemShut {NoStop}%
\bibitem [{\citenamefont {Fuchs}\ \emph {et~al.}(1998)\citenamefont {Fuchs},
  \citenamefont {G\"otze},\ and\ \citenamefont {Mayr}}]{fuchs1998asymptotic}%
  \BibitemOpen
  \bibfield  {author} {\bibinfo {author} {\bibfnamefont {M.}~\bibnamefont
  {Fuchs}}, \bibinfo {author} {\bibfnamefont {W.}~\bibnamefont {G\"otze}}, \
  and\ \bibinfo {author} {\bibfnamefont {M.~R.}\ \bibnamefont {Mayr}},\ }\href
  {\doibase 10.1103/PhysRevE.58.3384} {\bibfield  {journal} {\bibinfo
  {journal} {Phys. Rev. E}\ }\textbf {\bibinfo {volume} {58}},\ \bibinfo
  {pages} {3384} (\bibinfo {year} {1998})}\BibitemShut {NoStop}%
\bibitem [{\citenamefont {Voigtmann}(2011)}]{Voigtmann2011}%
  \BibitemOpen
  \bibfield  {author} {\bibinfo {author} {\bibfnamefont {T.}~\bibnamefont
  {Voigtmann}},\ }\href {\doibase 10.1209/0295-5075/96/36006} {\bibfield
  {journal} {\bibinfo  {journal} {Europhys. Lett.}\ }\textbf {\bibinfo {volume}
  {96}},\ \bibinfo {pages} {36006} (\bibinfo {year} {2011})}\BibitemShut
  {NoStop}%
\bibitem [{\citenamefont {Fuchs}\ \emph {et~al.}(1991)\citenamefont {Fuchs},
  \citenamefont {G\"otze}, \citenamefont {Hofacker},\ and\ \citenamefont
  {Latz}}]{Fuchs1991}%
  \BibitemOpen
  \bibfield  {author} {\bibinfo {author} {\bibfnamefont {M.}~\bibnamefont
  {Fuchs}}, \bibinfo {author} {\bibfnamefont {W.}~\bibnamefont {G\"otze}},
  \bibinfo {author} {\bibfnamefont {I.}~\bibnamefont {Hofacker}}, \ and\
  \bibinfo {author} {\bibfnamefont {A.}~\bibnamefont {Latz}},\ }\href
  {https://iopscience.iop.org/article/10.1088/0953-8984/3/26/022} {\bibfield
  {journal} {\bibinfo  {journal} {J. Phys.: Condens. Matter}\ }\textbf
  {\bibinfo {volume} {3}},\ \bibinfo {pages} {5047} (\bibinfo {year}
  {1991})}\BibitemShut {NoStop}%
\bibitem [{\citenamefont {Mandal}\ \emph {et~al.}(2019)\citenamefont {Mandal},
  \citenamefont {Schrack}, \citenamefont {L\"owen}, \citenamefont {Sperl},\
  and\ \citenamefont {Franosch}}]{Mandal2019}%
  \BibitemOpen
  \bibfield  {author} {\bibinfo {author} {\bibfnamefont {S.}~\bibnamefont
  {Mandal}}, \bibinfo {author} {\bibfnamefont {L.}~\bibnamefont {Schrack}},
  \bibinfo {author} {\bibfnamefont {H.}~\bibnamefont {L\"owen}}, \bibinfo
  {author} {\bibfnamefont {M.}~\bibnamefont {Sperl}}, \ and\ \bibinfo {author}
  {\bibfnamefont {T.}~\bibnamefont {Franosch}},\ }\href {\doibase
  10.1103/PhysRevLett.123.168001} {\bibfield  {journal} {\bibinfo  {journal}
  {Phys. Rev. Lett.}\ }\textbf {\bibinfo {volume} {123}},\ \bibinfo {pages}
  {168001} (\bibinfo {year} {2019})}\BibitemShut {NoStop}%
\end{thebibliography}
\end{document}